\documentstyle[12pt]{article}

\global\arraycolsep=1pt
\begin{document}

\hfill    SISSA/ISAS 49/94/EP

\hfill    IFUM 468/FT

\hfill    hepth@xxx/9404109

\hfill    April, 1994

\begin{center}
\vspace{10pt}
{\large \bf
CONSTRAINED TOPOLOGICAL GRAVITY\\
FROM TWISTED N=2 LIOUVILLE THEORY
\, \footnotemark
\footnotetext{Partially supported by EEC,
Science Project SC1$^*$-CT92-0789.}
}
\vspace{10pt}

{\sl Damiano Anselmi, Pietro Fr\'e}

\vspace{4pt}

International School for Advanced Studies (ISAS), via Beirut 2-4,
I-34100 Trieste, Italia\\
and Istituto Nazionale di Fisica Nucleare (INFN) - Sezione di Trieste,
Trieste, Italia\\

\vspace{8pt}

{\sl Luciano Girardello and Paolo Soriani}

\vspace{4pt}

Dipartimento di Fisica, Universit\`a di Milano, via Celoria 6,
I-20133 Milano, Italia\\
and Istituto Nazionale di Fisica Nucleare (INFN) - Sezione di Milano,
Milano, Italia\\
\end{center}

\vspace{8pt}

\begin{center}
{\bf Abstract}
\end{center}

\vspace{4pt}

\noindent
In this paper we show that there exists
a new class of topological field theories, whose
correlators are intersection numbers of cohomology classes in a constrained
moduli space. Our specific example is a formulation  of 2D topological gravity.
The constrained moduli-space is the Poincar\'e dual of the top Chern-class
of the bundle ${\cal E}_{hol}\longrightarrow {\cal M}_g$,
whose sections are the holomorphic differentials.
Its complex dimension is $2g-3$,
rather then $3g-3$.
We derive our model by performing the
A-topological twist of N=2 supergravity, that we identify with N=2 Liouville
theory, whose rheonomic construction is also
presented.
The peculiar field theoretical mechanism, rooted in BRST cohomology,
that is responsible for the constraint
on moduli space is discussed,
the key point being the fact
that the graviphoton becomes a Lagrange multiplier after twist.
The relation with conformal field theories is also explored.
Our formulation of N=2 Liouville theory leads to a representation of the
N=2 superconformal algebra with $c=6$, instead of the value $c=9$
that is obtained by untwisting the Verlinde and Verlinde formulation
of topological gravity.
The reduced central charge is the shadow, in conformal
field theory, of the constraint on moduli space.
Our representation of the N=2 algebra
can be split into the direct sum of a
minimal model with $c=3/2$ and a ``maximal''
model with $c=9/2$.
Considerations on the
matter coupling of constrained topological gravity are also presented.
Their study requires the analysis of both the A-twist and the  B-twist of
N=2 matter coupled supergravity, that we postpone to future work.
\vfill
\eject

\newcommand{\be}{\begin{equation}}
\newcommand{\ee}{\end{equation}}
\newcommand{\ba}{\begin{eqnarray}}
\newcommand{\ea}{\end{eqnarray}}
\newcommand{\ban}{\begin{eqnarray*}}
\newcommand{\ean}{\end{eqnarray*}}
\newcommand{\brr}{\begin{array}}
\newcommand{\err}{\end{array}}
\newcommand{\bc}{\begin{center}}
\newcommand{\ec}{\end{center}}
\newcommand{\sss}{\scriptscriptstyle}
\newcommand{\bea}{\begin{eqnarray}}
\newcommand{\eea}{\end{eqnarray}}
\newcommand{\bean}{\begin{eqnarray*}}
\newcommand{\eean}{\end{eqnarray*}}

\def\bz{{\bar z}}
\def\s{s^\prime}
\def\t#1{\tilde #1}
\def\La{\Lambda}
\def\lamb{\lambda}
\def\o#1#2{{{#1}\over{#2}}}
\def\nn{\nonumber}
\def\is{{I^*}}
\def\js{{J^*}}

\section{Introduction}
\label{intro}

The realm of topological field theories can be divided, according to Witten
\cite{witten}, in two broad classes:
the {\sl cohomological, or semiclassical theories}, whose
prototypes are either the
topological Yang-Mills theory \cite{wittenym} or the
topological $\sigma$-model \cite{wittensigma}
and the  {\sl quantum theories}, whose prototype is
the abelian Chern-Simons theory
\cite{chernsimons,report}.

In this paper we deal
with the cohomological theories and present new properties of these models.
In particular, we propose a new formulation of 2D topological
gravity \cite{witten,topgr},
leading to correlation functions apparently different
from those of Witten's theory. Nevertheless, our
considerations
have a more general scope and can apply to a wider
number of models.

The new features of cohomological theories that we analyse are the
reduction of moduli space to a constrained submanifold
of the ordinary moduli space and the field
theoretical mechanism that implements such a reduction.
Specifically,
we derive a theory of 2D topological gravity where
the physical correlators are
{\sl intersection numbers} in a proper submanifold ${\cal V}_{g,s}
\subset {\cal M}_{g,s}$
of the moduli space ${\cal M}_{g,s}$ of genus $g$ Riemann surfaces
$\Sigma_{g,s}$ with $s$ marked points.

${\cal V}_{g,s}$ is defined as follows.
Consider the $g$-dimensional vector bundle
${\cal E}_{hol} \, \longrightarrow \, {\cal M}_{g,s}$, whose
sections $s(m)$ are the holomorphic
differentials $\omega$ on the Riemann surfaces
$\Sigma_{g,s}$, $m$
denoting the point of the base-manifold ${\cal M}_{g,s}$
(i.e.\ the {\sl polarized} Riemann surface).
Let $c({\cal E}_{hol})=\det (1+{\cal R})$ be the total Chern class
of ${\cal E}_{hol}$, ${\cal R}$
being the curvature two-form of a holomorphic connection on
${\cal E}_{hol}$. For instance, we can choose the canonical connection
$\Gamma=h^{-1}\partial h$ associated with the natural fiber metric
$h_{jk}={\rm Im}\,
\Omega_{jk}$, $\Omega_{jk}$ being the period matrix of $\Sigma_{g}$.
Then ${\cal V}_{g,s}$ is the Poincar\'e dual
of the top Chern class $c_g({\cal E}_{hol})=\det {\cal R}=
\det\left({1\over \Omega-\bar\Omega}\bar\partial
\bar\Omega {1\over \Omega-\bar\Omega}\partial\Omega\right)$,
$d=\partial+\bar \partial$ being the exterior derivative on the moduli space.
${\cal V}_{g,s}$ is therefore a submanifold
of codimension $g$ described as the locus of
those Riemann surfaces $\Sigma_{g,s}(m)$ where
some section $s(m)$ of ${\cal E}_{hol}$ vanishes \cite{griffithsharris}.

Explicitly, the
topological correlators of our theory
are the intersection numbers of the standard Mumford-Morita
cohomology classes $ c_1 \left ( {\cal L}_i \right )$ on the constrained
moduli space, namely
\begin{eqnarray}
<{\cal O}_1 \left ( x_1 \right )  {\cal O}_2 \left ( x_2 \right )
&\cdots &{\cal O}_n \left ( x_n \right )  > \, =\int_{{\cal V}_{g,s}}
\left [  c_1  \left ( {\cal L}_1 \right ) \right ]^{d_1}  \wedge  \cdots\wedge
\left [  c_1  \left ( {\cal L}_n \right ) \right ]^{d_n}=\nonumber\\&
=&\int_{{{\cal M}}_{g,s}} c_g (  {\cal E}_{hol})\wedge
\left [  c_1 \left ( {\cal L}_1 \right ) \right ]^{d_1}  \wedge \cdots
\wedge\left [  c_1 \left ( {\cal L}_n \right ) \right ]^{d_n}.
\label{intro_0}
\end{eqnarray}
Precisely,
$ c_1 \left ( {\cal L}_i \right )$ are the first Chern-classes
of the line bundles
${\cal L}_i\longrightarrow{\cal M}_{g,s}$ defined
by the cotangent spaces
$T^{ * }_{x_i} \Sigma_{g}(m)$ at the marked points $x_i$.
The above theory will be called {\sl 2D constrained topological gravity}.

We derive the definition of 2D constrained
topological gravity in an algorithmic
way, by performing
the topological twist of N=2 Liouville theory, that we assume as the
correct definition of
N=2 supergravity in two-dimensions. The formal set-up for
twisting an N=2 locally supersymmetric theory was established in ref.\
\cite{ansfre}.

The origin of a constraint on moduli space is due to the presence
of the graviphoton, absent in the existing
formulations of 2D topological
gravity. The graviphoton is initially a physical gauge-field and
after the twist it maintains zero ghost-number. Nevertheless,
in the twisted theory, it is no longer a physical field,
rather it is a Lagrange multiplier (in the BRST sense). Indeed,
it appears in the right-hand side of the BRST-variation of suitable
antighosts. Since this Lagrange multiplier possesses global degrees
of freedom (the $g$ moduli of the graviphoton), it
imposes $g$ constraints on the space ${\cal M}_{g}$, which can be
viewed as the space
of the global degrees of freedom of the metric tensor. The metric
tensor,
on the other hand, is the only
field that remains physical also after twist.
We are lead to conjecture that
the inclusion of {Lagrange multiplier gauge-fields}
is a general mechanism producing the appearance of {constrained}
moduli spaces. We recall that
in four dimensions, instead, the role of the graviphoton
$A$ \cite{ansfre,ansfre2} is that of producing ghosts for ghosts {\sl via}
the self-dual and anti-self-dual components $F^{\pm \, ab}$ of the field
strength $F^{ab}$.

To develop further the introductory description of our theory, we begin
with a brief discussion of cohomological theories from our
specific viewpoint.

In Witten's words \cite{witten},
cohomological field theories are concerned with
{\sl sophisticated counting
problems}. The fundamental idea is that
a generic correlation function of $n$ physical observables
$\{ {\cal O}_1 , \ldots, {\cal O}_n \}$
has an interpretation as the {\sl intersection number}
\begin{equation}
<{\cal O}_1  {\cal O}_2 \cdots {\cal O}_n > \,=\,\# \left (  H_1  \cap  H_2
 \cap  \cdots \cap H_n  \right )
\label{intro1}
\end{equation}
of $n$ {\sl homology cycles} $H_i \, \subset \, {\cal M}$ in the {\sl moduli
space} ${\cal M}$
of suitable {\sl instanton} configurations $\Im \left [\phi (x) \right ]$ of
the basic
fields $\phi$ of the theory.
For example in the topological $\sigma$-model
\cite{wittensigma,topsigma,calabi,index1,index2}
the basic fields are the maps
\begin{equation}
X:\, \Sigma_g \longrightarrow{\cal N}
\label{intro2}
\end{equation}
from a genus $g$ Riemann surface $\Sigma_g$ into a K\"ahlerian
manifold ${\cal N}$. In this case
the instantons $\Im \left [X (z,{\bar z}) \right ]$ are the holomorphic
maps
$\partial_{\bar z} X\, = \,\partial_{z}{\bar X }\, = \, 0$ and the moduli
space is, for each
homotopy class of holomorphic embeddings of degree $k$, the parameter
space ${\cal M}_k$ of such class
of maps. The degree $k$ is defined by $\int_{\Sigma_g}  X^{ * } K =k$,
where $X^{ * } K$
is the pull-back of the K\"ahler two-form $K$
on ${\cal N}$. The observables
${\cal O}_{A_i}(z_i)$ are in
one-to-one correspondence with the de Rahm cohomology classes
$A_i  \in H^{p} ({\cal N})$ of
the target manifold ${\cal N}$  and the homology
cycles $H_i$ are defined
as the subvarieties of ${\cal M}_k$ that contain all those instantons
such that
$X(z_i) \in   [ A_i ]^{ * }$.
In this definition we have denoted by $[ A_i ]^{ * }
\subset
{\cal N }$ the Poincar\'e dual of the cohomology class $A_i  \in
H^{p}({\cal N})$.

It is clear that
topological field theories \cite{report}
can been defined in completely geometrical terms.
However, in every topological model,
the right hand side of equation (\ref{intro1}) should admit an
independent definition as a
functional integral in a suitable Lagrangian quantum field theory,
in order to be of physical interest.
The basic feature of the classical Lagrangian
is that of possessing
a very large group of gauge symmetries,
the {\sl topological symmetry}, which is the most general continuous
deformation
of the classical fields. The topological symmetry is treated
through the standard techniques of BRST quantization and
the instanton equations are imposed as a gauge-fixing.
In this
way, eq. (\ref{intro1}), rather than a definition, becomes a map
between a {\sl physical} and
a {\sl mathematical} problem, which evenience is the main source
of interest for topological field theories.

{}From the physical point of view,  the basic
properties of a topological field
theory are encoded in the BRST algebra ${\cal B}$ \cite{BRST,bonora} and
the anomaly of ghost number.

The moduli space cohomology originates the cohomology
of the BRST operator $s$:
the left-hand side of eq. (\ref{intro1}) is
the vacuum expectation
value of the product of $n$ representatives ${\cal O}_i$
of  non-trivial BRST cohomology classes
\begin{equation}
s {\cal O}_i =0,\quad\quad
{\cal O}_i \ne  s\{{\rm anything} \}.
\label{intro13}
\end{equation}
Correspondingly, the right-hand side of eq.\ (\ref{intro1}) can be
expressed as an integral of a product of cocycles over the moduli space.

In full generality, the BRST algebra ${\cal B}$ can be decomposed as
\begin{equation}
{\cal B}={\cal B}_{gauge-free}\oplus{\cal B}_{gauge-fixing},
\label{intro3}
\end{equation}
where ${\cal B}_{gauge-free} \subset {\cal B}$ is the subalgebra
that contains only the
physical fields and the ghosts (fields of non negative ghost number),
while ${\cal B}_{gauge-fixing}$ is the extension
of ${\cal B}_{gauge-free}$ by means of antighosts and Lagrange
multipliers (or the corresponding gauge-fixing conditions),
of non positive ghost number. Usually, ${\cal B}_{gauge-fixing}$
is trivial, but this is not the case we deal with, since the
interesting features of our theory come precisely from the
nontrivial nature of ${\cal B}_{gauge-fixing}$.

We postpone the discussion of the structure of ${\cal B}_{gauge-free}$
and ${\cal B}_{gauge-fixing}$, in order to consider the
mathematical meaning of the other basic
aspect of the field theoretical approach, i.e.\ the anomaly of
ghost number.
The left-hand side of eq. (\ref{intro1}) can be non-zero only if
\begin{equation}
\sum_{i}  d_i =\Delta U =\int  \partial^\mu J^{(ghost)}_\mu d^D x,
\label{intro13b}
\end{equation}
where $J^{(ghost)}_\mu $ is the ghost-number current, $\Delta U$ is its
integrated anomaly and $d_i= gh[{\cal O}_i]$ is the ghost number of
${\cal O}_i$ \cite{index1,index2}.
The divergence of the ghost-current has an interpretation as index-density
for some elliptic
operator $\nabla$ that appears in the quantum action through the kinetic
term of the ghost($C$)-antighost($\bar C$)
system:
\begin{equation}
S_{quantum}=\int  ( \cdots + {\bar C }  \nabla  C  + \cdots ).
\label{intro14}
\end{equation}
We have
\begin{equation}
{\rm index}_{\nabla}=\# ~{\rm zero~modes~of~ghosts} ~-~\#~{\rm zero~modes~of~
antighosts}.
\label{intro15}
\end{equation}
On the other hand, the right-hand side of eq. (\ref{intro1}) can be
non-zero only if the sum of the codimensions of the homology
cycles $H_i$
adds up to the total dimension of the moduli space,
\begin{equation}
\sum_{i}  {\rm codim} \, H_i ~=~{\rm dim} \, {\cal M}.
\label{intro15b}
\end{equation}
In other words, the physical observables must reduce, after functional
integration on the irrelevant
degrees of freedom, to cocycle forms $\Omega_i$ of degree $d_i$ on
the moduli-space
${\cal M}$ (the Poincar\'e duals of the cycles $H_i$) and their wedge
product must be a top-form.
This means
\begin{equation}
\Delta U={\rm dim} \,{\cal M}.
\label{intro16}
\end{equation}
Such an equation is understood in the following way. In the
background of an instanton,
namely of a gauge-fixed configuration, the zero-modes of the
topological ghosts
correspond to the
residual infinitesimal deformations that  preserve the gauge condition.
Their number is therefore
the dimension of the tangent space to the parameter space of the instanton.
The zero-modes
of the antighosts correspond, instead, to potential global obstructions to
the integration
of these infinitesimal deformations \cite{index1,index2}. The index $\Delta U$
is therefore named the formal dimension
of the moduli space ${\cal M}$. The true dimension  of
the moduli space
is larger or equal to its formal dimension,
\begin{equation}
{\rm dim}^{true}{\cal M}\ge{\rm dim}^{formal}{\cal M}=\Delta U,
\label{intro17}
\end{equation}
depending on whether the potential obstructions become real obstructions
or not.

In the case of 2D topological gravity, Witten started \cite{witten}
from the right-hand
side of eq. (\ref{intro1}),
proposing a completely geometrical definition. He
assumed that the
relevant moduli-space is the standard moduli-space ${\cal M}_{g,s}$
of Riemann surfaces of
genus $g$ with $s$ marked points, whose dimension is well known to be
\begin{equation}
{\rm dim}_{\bf C} {\cal M}_{g,s}=3g  -  3  +  s
\label{intro18}
\end{equation}
and identified the observables ${\cal O}_i$
with the Mumford-Morita cohomology classes, namely the
$2d_i$-forms
$\left [  c_1 \left ( {\cal L}_i \right ) \right ]^{d_i}$ \cite{moritamumford}
on ${\cal M}_{g,s}$  introduced in eq. (\ref{intro_0}).
Correspondingly, Witten obtained the selection rule:
\begin{equation}
\sum_{i=1}^{s} d_i =3g -  3  +  s.
\end{equation}
For a reason that will be clear in a moment, it is convenient, from
the
field theoretical point of view, to rewrite this condition as
\begin{equation}
\sum_{i=1}^{s}  ( d_i -  1) =3g  -  3,
\label{intro19}
\end{equation}
where now the right hand side is the dimension of the moduli space
${\cal M}_{g}$ without marked points.
In this way, Witten assumed that in the field-theoretical formulation
of 2D topological
gravity, whatever it might be, the integrated anomaly of the
ghost-number current should be
\begin{equation}
\Delta U =\int  \partial^\alpha J^{(ghost)}_\alpha d^2 x=
3g  -  3.
\label{intro20}
\end{equation}
To understand this way of formulating the sum rule, we have to recall
the concept of
descent equations for the physical observables.
Every local observable ${\cal O}_i$ of 2D topological gravity can be
written as $\sigma_{d_i}^{(0)}(x_i)=\gamma_0^{d_i}(x_i)$,
$\gamma_0(x)$ being a suitable composite field.
${\cal O}_i$ is a zero form of ghost-number
$2 d_i$ and it is related to a
one-form $\sigma_{d_i}^{(1)}(x_i)$ of ghost number $2d_i-1$ and to
a two-form
$\sigma_{d_i}^{(2)}(x_i)$ of ghost number $2(d_i-1)$
{\sl via} the descent equations
\begin{equation}
s\sigma_{d_i}^{(0)}=0,\quad\quad
s \sigma_{d_i}^{(1)}= d \sigma_{d_i}^{(0)},\quad\quad
s \sigma_{d_i}^{(2)}=d \sigma_{d_i}^{(1)},\quad\quad 0=d\sigma^{(2)}_{d_i}.
\label{intro21}
\end{equation}
As a consequence, the integrated observables
$\int_{\Sigma_{g}}  \sigma_{d_i}^{(2)}$ are BRST-closed,
\begin{equation}
s  \int_{\Sigma_{g}}  \sigma_{d_i}^{(2)}=0,
\label{intro22}
\end{equation}
and can be traded for the local ones, by an equivalence
\begin{equation}
< {\cal P}(x_1)\sigma_{d_1}^{(0)}(x_1) \cdots
{\cal P}(x_n)\sigma_{d_n}^{(0)}(x_n) >\,\approx\,
< \int_{\Sigma_{g}} \sigma_{d_1}^{(2)}(x_1) \cdots
\int_{\Sigma_{g}} \sigma_{d_n}^{(2)}(x_n) >.
\label{intro23}
\end{equation}
${\cal P}(x_i)$ denote certain {\sl picture changing operators}
\cite{verlindesquare}, that
have ghost number $-2$ and whose mathematical meaning is that
of marking the points $x_i$ where the local operators $\sigma^{(0)}(x_i)$
are inserted.
Both members of (\ref{intro23}) can be calculated as intersection
integrals over ${\cal
M}_g$. The integrations appearing on the right hand side, however, can be
also understood as integrations over the positions of ``marked
points'' $x_i$, so that one is allowed to conjecture the
correspondence
\begin{equation}
\sigma^{(2)}_{d_i}(x_i)\sim \gamma_0^{d_i}(x_i)\sim
[c_1({\cal L}_i)]^{d_i},
\quad \gamma_0(x_i)\sim c_1({\cal L}_i).
\end{equation}

Eq.\ (\ref{intro23}) says that the topological amplitude can also be viewed
as the correlator of BRST cohomology classes of degree $2(d_i-1)$
on the space ${\cal M}_{g}$ and
Witten's conjecture (\ref{intro20}) on the integrated anomaly
of the ghost-current in any field-theoretical
formulation of the theory is explained.
Indeed,  in \cite{verlindesquare} Verlinde and Verlinde constructed
an explicit field theory model where eq. (\ref{intro20}) is verified.

On the contrary, the key result of the present paper is the following one.
We present a different field
theoretical model of topological gravity where eq.(\ref{intro20}) is
replaced by
\begin{equation}
\Delta U =\int  \partial^\alpha  J^{(ghost)}_\alpha d^2 x=
2g  -  2
\label{intro21b}
\end{equation}
The geometrical interpretation of this fact has already been anticipated.
Eq. (\ref{intro21b})
indicates that we are dealing with a {\sl constrained} moduli space
${\cal V}_{g}$
whose formal complex dimension is
${\rm dim}^{formal}_{\bf C}\, {\cal V}_{g} =2 g -  2$. Actually
the constrained
moduli-space ${\cal V}_{g}$
is the Poincar\'e dual of $c_g({\cal E}_{hol})$ and
its true dimension turns out to be
${\rm dim}^{true}_{\bf C}{\cal V}_{g} =2g-3$,
which is smaller than
the formal dimension, another apparently puzzling result. However,
if we recall that the effective
moduli space emerges from a constraint on a larger moduli-space,
then the fact that the formal-dimension
is bigger than the actual dimension becomes less mysterious. Indeed,
this time, in the sector of
the BRST-algebra that implements the constraint, usual rules are inverted.
Antighost zero-modes correspond
to local vector fields normal to the constrained surface and ghost
zero-modes correspond to possible obstructions to the
globalization of such local vector fields. As a consequence, the difference,
in the constraint sector of the BRST algebra, of antighost zero-modes
minus ghost zero-modes,  expresses the minimum number of
constraints that are imposed. If the potential obstructions do not occur,
then all the
antighosts correspond to actual normal directions to the constrained
surface and the true
dimension of the constraint surface is smaller than its formal dimension.

Notice that the constraint imposed  is not a ``generic'' constraint,
but a BRST constraint, i.e.\ a gauge-fixing of some symmetry.
This translates geometrically into the fact, already pointed out, that the
constrained moduli space is not a specific hypersurface in ${\cal M}_g$,
rather it is a ``slice choice''
of a representative in a homology class of closed
submanifolds (the Poincar\'e dual of $c_g({\cal E}_{hol})$).

Let us now discuss the general structure of our topological field-theory
in comparison
with that introduced by Verlinde and Verlinde in \cite{verlindesquare}.
The basic idea of \cite{verlindesquare} is
that the moduli-space of Riemann surfaces $\Sigma_g$ can be related
to the moduli-space of
$SL(2,R)$ flat connections on the same surface. This goes back to the
classical Fenchel-Nielsen
parametrization of the Teichmuller space. A flat $SL(2,R)$ connection
$\left \{ e^\pm , e^0 \right \}$
contains the zweibein $e^\pm$
and the spin connection of a constant curvature metric on the imaginary
upper half-plane $H$.
If that connection is pull-backed to the quotient $H/\Gamma_g$, where
$\Gamma_g$ is a Fuchsian
group realizing the homotopy group $\pi_1 \left ( \Sigma_g \right )$ of
a genus  $g$
surface, then the connection $\left \{ e^\pm , e^0 \right \}$ realizes a
constant curvature
metric on that surface. In view of this, the authors of \cite{verlindesquare}
identified
the gauge-free BRST  algebra ${\cal B}_{gauge-free}^{2D~grav}$ of 2D
topological gravity with
${\cal B}_{gauge-free}^{\bf SL(2,R)}$, namely the {\sl gauge-free}
topological algebra
associated with the Lie-algebra ${\bf SL(2,R)}$.
For a Lie algebra ${\bf g}$ with structure constants $f^I_{\phantom{I}JK}$,
$I=1,\ldots {\rm dim}\, {\bf g}$,
${\cal B}_{gauge-free}$ is
\begin{eqnarray}
s A^I&=& \Psi^I-  dC^I-f^I_{\phantom{I}JK}A^JC^K, \nonumber\\
s \Psi^I &=& -d\Gamma^I- f^I_{\phantom{I}JK}A^J\Gamma^K-f^I_{\phantom{I}JK}
C^J\Psi^K, \nonumber\\
s  C^I&=&\Gamma^I-{1\over 2}f^I_{\phantom{I}JK}C^JC^K,\nonumber\\
s  \Gamma^I &=&  f^I_{\phantom{I}JK}C^J\Gamma^K,
\end{eqnarray}
$\Psi$ being the topological ghost, $C$ the ordinary gauge ghost
and $\Gamma$ being the ghost for the ghosts.
In the case of ${\bf SL(2,R)}$, the structure constants $f^I_{\phantom{I}JK}$
are encoded in the curvature definitions
\begin{equation}
R^\pm = d  e^{\pm}  \pm  e^0  \wedge \, e^{\pm},\quad\quad
R^0 = d  e^0  +  a^2 e^+  \wedge e^-,
\end{equation}
where $a^2\in{\bf R}_+$ expresses the size of the
constant negative
curvature.
The  gauge-fixing algebra
${\cal B}_{gauge-fixing}^{\bf SL(2,R)}$ introduced by Verlinde and
Verlinde realizes in an obvious
manner the geometrical idea of flat-connections. The primary
topological symmetry is broken
by introducing antighosts whose BRST-variations are the Lagrange
multipliers for the
constraints $R^\pm = R^0=0$. In addition antighosts and Lagrange
multiplier are also
introduced to fix diffeomorphisms, Lorentz invariance
and the gauge-symmetry of the topological ghosts,
namely superdiffeomorphisms.
After gauge-fixing and in the limit
$a^2\rightarrow 0$, the model of ref.\ \cite{verlindesquare}
reduces to the sum of two topological
conformal field theories $Liouville\, \oplus \, Ghost$, that can be
untwisted to N=2 conformal
field-theories of central charges $c_{Liouville}=9$ and  $c_{Ghost}=-9$.
\par
As one sees, this construction is in the spirit of the Baulieu-Singer
approach to topological
field-theories, where the gauge-fixing sector is invented {\sl ad hoc}.
On the other hand,
the topological twisting algorithm produces topological theories
where the gauge-fixing part of the BRST-algebra is already encoded in
the original untwisted
N=2 field-theory model.  For instance, applying these ideas to the case
of D=4, N=2
$\sigma$-models we discovered the concept of hyperinstantons
\cite{ansfre2}.
In this paper we adopt the
twisting strategy also in the case of 2D topological gravity. The necessary
input is a
definition of two-dimensional N=2 supergravity. This problem is solved by
N=2 supersymmetrizing
a reasonable definition of two-dimensional gravity. Following
\cite{cange,teitel} we assume as Lagrangian of
ordinary 2D gravity
the following one:
\begin{equation}
{\cal L}_{2D-grav}=\Phi(R[g]+a^2) \sqrt { \det g}
\label{intro923}
\end{equation}
where $\Phi$  and the metric $g_{\alpha\beta}$ are treated as
independent fields.
The variation in $\Phi$ imposes the constant curvature constraint
on $g_{\alpha\beta}$.
The Lagrangian (\ref{intro923}) is equivalent,
through the
field redefinition $g_{\mu\nu}\rightarrow g_{\mu\nu}{\rm e}^\Phi$
to the more conventional Liouville Lagrangian
\begin{equation}
{\cal L}_{Liouville} =[\nabla_\alpha \Phi \nabla^\alpha \Phi +
\Phi(R[g]+a^2  {\rm e}^\Phi ) ]
\sqrt { \det  g}.
\label{intro24}
\end{equation}
Both Lagrangians (\ref{intro923}) and (\ref{intro24}) can be N=2
supersymmetrized and the results
are related to each other by a field redefiniton as in the N=0 case.
Hence, we work with the
N=2 analogue of the simpler form (\ref{intro923}) and to it we apply
the topological twist.
The rest follows, although it requires interpretations that are by no
means straightforward. As anticipated, the
essential new feature is the presence, in the N=2 gravitational
multiplet, of the
graviphoton,  a $U(1)$ gauge-connection $A$ that maintains
ghost number $0$ after the twist.  Hence,
the geometrical structure we deal with is that of a $U(1)$ bundle
on a Riemann surface.
The possible deformations of this structure are more than the
deformations of  a bare
Riemann surface. Indeed, the total number of moduli for this
bundle is $4g - 3$,
$g$ new moduli being contributed by the deformations of ${A}$.
The naive conclusion
would be that the gauge-free topological algebra underlying the
twisted theory is the
direct sum ${\cal B}_{gauge-free}^{\bf SL(2,R)} \oplus
{\cal B}_{gauge-free}^{\bf U(1)}$.
This would lead to intersection theory in the $4g-3$
dimensional moduli-space
of the $U(1)$-bundle over $\Sigma_g$, but it is not the case. What
actually happens is that
the graviphoton belongs to ${\cal B}_{gauge-fixing}$, rather than
to ${\cal B}_{gauge-free}$,
satisfying a BRST-algebra of the type
\begin{equation}
s{\bar\psi}=A-d{\gamma},\quad\quad sA=-dc,\quad\quad s\gamma=c,\quad
\quad sc=0,
\label{intro25}
\end{equation}
where ${\bar \psi}$ is a one-form of ghost number $-1$,
${\gamma}$ is a zero-form of ghost number $0$ and
$c$ is the ordinary gauge ghost (with ghost-number $1$).
The geometrical meaning of the gauge-fixing
algebra (\ref{intro25}) has already been anticipated. The deformations
of $A$ correspond to constraints on the allowed deformations of the bundle
base-manifold, namely of the Riemann surface.
How this mechanism is implemented by the functional integral
is what we show in the later technical sections of our paper.
We conjecture that BRST-algebras of type (\ref{intro25}), associated
with a principal G-bundle $E \rightarrow M$  always correspond
to constraints on the moduli-space of the base-manifold $M$.

Hence, our formulation of 2D topological gravity is based on the same
{\sl gauge-free}
algebra ${\cal B}_{gauge-free}^{\bf SL(2,R)}$ as the model of Verlinde
and Verlinde, but it has a much different and more subtle
{\sl gauge-fixing} algebra. At the level of conformal
field theories there is also a crucial difference, which
keeps trace of the constraint on moduli space.
Indeed, after gauge-fixing
and in the limit $a^2\rightarrow 0$, our model also reduces to the sum
of two topological
conformal field theories $Liouville\, \oplus \, Ghost$;
the central charges, however, are   $c_{Liouville}=6$
and  $c_{Ghost}=-6$,
rather than $9$ and $-9$. We discuss the structure of these conformal
field theories in more detail later on.

Let us conclude by summarizing our
viewpoint. Equation (\ref{intro1}) possesses two deeply different
meanings,
depending on whether one reads it from the left (= physics)
to the right (= mathematics), or from the right to the left.

i) From the right to the left, equation (\ref{intro1})  means:

``given a well defined mathematical problem (intersection theory on the
moduli space of the instantons of some class of maps), find a quantum
field theory
 that represents the intersection forms as physical amplitudes (averages
 of products of physical observables)''.

This problem is solved by BRST quantizing the most general continuous
deformations
of the classical fields (which sort of fields depending on the class of maps
that is under consideration) and imposing the instantonic equations as
a gauge-fixing.

ii) From the left to the right, (\ref{intro1}) means

``given a physically well-defined topological quantum field theory
(as it is the topological twist of an N=2 supersymmetric theory), find
the mathematical problem (maps, instantons, intersection theory)
that it represents.

Of course, there is no general recipe for solving this second problem.
We started considering it in four dimensions \cite{ansfre,ansfre2,ansfre3}
and this lead to interesting results. The topological twist
of the N=2 supersymmetric $\sigma$-model permitted to
introduce a concept of instantons ({\sl hyperinstantons})
\cite{ansfre2} that we later
\cite{ansfre3} identified with a triholomorphicity condition on the
embeddings of four dimensional almost quaternionic manifolds into
almost quaternionic manifolds. With this paper, we continue the program
begun in ref.\ \cite{billofre} of considering the same problem
in two dimensions, aiming to uncover the mathematical
meanings of the topological field theories obtained by twisting the D=2
N=2 supersymmetric theories. We feel that the amazing secrets
of N=2 supersymmetry have not yet been fully uncovered.

Our paper is organized in three main parts. The first part (sections
\ref{N2D2}-\ref{twist})
is devoted to the
construction of N=2 Liouville theory and its twist. In section
\ref{N2D2} we present D=2 N=2
supergravity in the rheonomic approach,
while in section \ref{coupling} we couple it to N=2
Landau-Ginzburg chiral matter \cite{LG,topLG}. In section \ref{liouville},
we derive the Lagrangian of N=2 Liouville theory, that involves
a suitable combination of the gravitational multiplet
and a simple chiral multiplet.
We BRST quantize the theory in section \ref{gaugefree} and perform the
topological twist in section \ref{twist}.
The second part (sections \ref{conformal} and \ref{twist1}) is devoted
to conformal field theory.
In section \ref{conformal}, we gauge-fix N=2 Poincar\`e gravity and show
that it corresponds to a conformal field theory
with $c=c_{Liouville}+c_{ghost}$,
$c_{Liouville}=6$ and $c_{ghost}=-6$. We study the
N=2 currents, the BRST current and various conformal properties.
In section \ref{twist1} we study the topological twist of the
gauge-fixed theory, describing the match between the general twist procedure
of section \ref{twist} and the procedure that is well known
in conformal field theories \cite{eguchiyang}.
We study some properties that emphasize
the differences between our model and the Verlinde and Verlinde one.
The third part corresponds to section \ref{geometry}, where
we suggest the mathematical interpretation of the
topological theory and draw
the correspondence between quantum field theory and geometry.
Finally in section \ref{concl} we address some open problems.

\section{N=2 D=2 supergravity}
\label{N2D2}

Following \cite{cange} we assume as the classical Lagrangian of pure gravity
the one displayed in eq.\ (\ref{intro923}).
Alternatively, in view of the equivalence
between (\ref{intro923}) and (\ref{intro24}),
we can also describe the action of
pure 2D gravity as the Polyakov action for a Liouville system.
We insist on the concept of Polyakov formulation, since the key point
in eq.\ (\ref{intro923}) is that both $\Phi$ ad $g_{\alpha\beta}$ have
to be treated as independent fields. This being clarified, we define
N=2, D=2 supergravity as the supersymmetrization of eq.\ (\ref{intro923}).
To perform such a supersymmetrization, we need the following
two ingredients.

i) An off-shell representation of the N=2 algebra, that corresponds to the
graviton multiplet containing the metric $g_{\alpha\beta}$.

ii) An off-shell representation of the N=2 algebra that corresponds
to the chiral scalar multiplet containing the field $\Phi$.

The final Lagrangian is obtained by combining
these two multiplets.

For the purpose of this construction, we use the so called rheonomic
formalism. This formalism was
originally proposed in 1978-79
\cite{libro}, as an alternative method  with respect to
the superfield formalism in studying
supersymmetric theories.
Its main advantage is the reduction of the
computational effort for constructing supersymmetric
theories to a simple series of geometrically meaningful steps.
{\sl Rheonomy} means ``law of the flux'' and refers to the
fact that the supersymmetrization of a theory can be viewed as a
Cauchy problem, in which spacetime, described by the inner (bosonic)
components $x$ of superspace, represents the boundary, while the outer
(Grassmann) components $\theta$
represent the direction of ``motion'': the rheonomic
principle represents the ``equation of motion''.
At the end of the rheonomic procedure, the only free choice
is the boundary condition, which is the spacetime theory,
projection of the superspace theory onto the inner components.
The fields (or
more generally differential forms) are functions on superspace and
the supersymmetry transformations are viewed as odd diffeormorphisms in
superspace.
We shall give, along the paper, many details on the rheonomic approach
in order to furnish enough information for using it.

In a recent paper \cite{billofre},
the rheonomic formalism was applied to the construction
of D=2 globally supersymmetric N=2 systems. In particular,
the rheonomic formulation of N=2 chiral multiplets
coupled to N=2 gauge multiplets
was provided. This involved the solution of Bianchi identities
and the construction of rheonomic actions in the background of a flat
N=2 superspace.

In this section we generalize the construction to curved superspace.
To this effect an important preliminary point to be discussed is the
following. The formulation of supergravity is performed in the
physical Minkowski signature $\eta_{ab}={\rm diag} (+,-)$, yet, after
Wick rotation we shall deal with supersymmetry defined on compact
Riemann surfaces. These have a positive, null or negative
curvature ${\cal R}$ depending on their genus $g$, (${\cal
R} > 0$ for $g=0$, ${\cal R}=0$ for $g=1$ and ${\cal R}<0$ for
$g \ge 2$). The sign of the curvature in the Euclidean theory is
inherited from the sign of the curvature in the Minkowskian formulation.
Since we want to discuss the theory for all genera $g$, we need
formulations of supergravity that can accommodate both signs of the
curvature. From the group-theoretical point of view, curved superspace
with ${\cal R}> 0$ is a continuous deformation of the supersymmetric
version of the de Sitter space, whose isometry group is $SO(1,D)$, $D$
being the space-time dimension. Hence, for ${\cal R}>0$ the
appropriate superalgebra to begin with is, if it exists, the N-extended
supersymmetrization of $SO(1,D)$. Similarly, for ${\cal R}<0$,
curved superspace is a continuous deformation of the supersymmetric
version of the anti de Sitter space, whose isometry group is $SO(2,D-1)$.
Hence, in this second case, the appropriate superalgebra to start from
in the construction of supergarvity is the N-extended
supersymmetrization of $SO(2,D-1)$. Alternatively, one can start from
the Poincar\'e superalgebra, that corresponds to an Inon\"u-Wigner
contraction of either the de Sitter or the anti de Sitter algebra,
and reobtain either one of the decontracted algebras as
vacua configurations alternative to the Minkowski one,
by giving suitable expectation values to the auxiliary
fields appearing in the rheonomic parametrizations of the Poincar\'e
curvatures. Actually, in space-time dimensions different from D=2, the
supersymmetric extensions of both the de Sitter and the anti de Sitter
algebras are not guaranteed to exist. For instance, in the relevant D=4
case, the real orthosymplectic algebra $Osp(4/N)$ is the
N-superextension of the anti de Sitter algebra $SO(2,3)$ but a
superextension of the de Sitter algebra $SO(1,4)$ does not exist.
This is the group-theoretical rationale of some otherwise well known facts.
In four-dimensions all supergravity vacua with a positive sign of
the cosmological constant (de Sitter vacua) break supersymmetry
spontaneously, while the only possible supersymmetric vacua are either
in Minkowski or in anti de Sitter space. Indeed, starting from a
formulation of D=4 supergravity based on the Poincar\'e superalgebra,
one obtains both de Sitter and anti de Sitter vacuum configurations
through suitable expectation values of the auxiliary fields,
but it is only in the
anti de Sitter case that these  constant expectation values are
compatible with the Bianchi identities of a superalgebra, namely
respect supersymmetry \cite{libro}.
In the de Sitter case the gravitino develops
a mass and supersymmetry is broken. In other words, in D=4,
supersymmetry chooses a definite sign for the curvature ${\cal R}<0$.
\par
If this were the case also in $D=2$, supersymmetric theories could not
be constructed on all Riemann surfaces, but only either in genus $g\ge
1$ or in genus $g\leq 1$. Fortunately, for D=2 it happens that the de
Sitter group $SO(1,2)$ and the anti de Sitter group $SO(2,1)$ are
isomorphic. Hence, a supersymmetrization of one is also a
supersymmetrization of the other, upon a suitable correspondence.
Once we have fixed the conventions for what we
call the physical zweibein, spin connection and gravitini, we obtain
an off-shell formulation of supergravity where the sign of the
curvature is fixed: it is either non-negative or non-positive.
Through
a field correspondence,
we can however make a transition from one case to
the other, but a continuous deformation of the auxiliary field
vacuum expectation value is not sufficient for this
purpose. Furthermore, as we are going to see, the Inon\"u-
Wigner contraction of the N=2 algebra displays also some new features
with respect to the $U(1)$ generator associated with the graviphoton.

The most general $D=2$ superalgebra one can write down,
through Maurer-Cartan equations, is obtained by setting to zero the
following curvatures:
\ba
T^+ &=& de^+ + \omega e^+ -\o{i}{2} \zeta^+ \zeta^-, \nn\\
T^- &=& de^- - \omega e^- -\varepsilon\o{i}{2} \tilde\zeta_+ \tilde\zeta_-,
\nn\\
\rho^+ &=& d \zeta^+ + \o{1}{2} \omega \zeta^+ +\o{ia_1}{4} A\zeta^+
+a_1 a_2 \tilde\zeta_- e^+, \nn\\
\rho^- &=& d \zeta^- + \o{1}{2} \omega \zeta^- -\o{ia_1}{4} A\zeta^-
+a_1 a_2 \tilde\zeta_+ e^+, \nn\\
\tilde\rho_+ &=& d \tilde\zeta_+ -\o{1}{2} \omega \tilde\zeta_+ -\o{ia_1}{4} A
\tilde\zeta_+ -\varepsilon a_1 a_2 \zeta^- e^-,\nn\\
\tilde\rho_- &=& d \tilde\zeta_- -\o{1}{2} \omega \tilde\zeta_- +\o{ia_1}{4} A
\tilde\zeta_-
-\varepsilon a_1 a_2 \zeta^+ e^-,\nn\\
R &=& d \omega -2 \varepsilon a_1^2 a_2^2 e^+ e^- -\o{i}{2} a_1 a_2 (\zeta^+
\tilde\zeta_+ + \zeta^- \tilde\zeta_-),\nn\\
F&=& dA -a_2 (\zeta^- \tilde\zeta_- -\zeta^+ \tilde\zeta_+),
\label{rhbi}
\ea
where $e^+$ and $e^-$ denote the two components (left and right
moving) of the world sheet zweibein one form, while $\zeta^+$,
$\tilde\zeta_+$ are the two
components of the gravitino one form, $\zeta^-$,
$\tilde\zeta_-$ are the two components of its complex conjugate.
$\varepsilon$ can take the values $\pm 1$ and distinguishes the de
Sitter ($\varepsilon =1$) and anti de Sitter ($\varepsilon =-1$)
cases. Formally, one can pass from positive to negative curvature
by replacing $e^-$ and $T^-$ with $-e^-$ and $-T^-$.

The algebra (\ref{rhbi}) contains two free (real) parameters $a_1, a_2$ in its
structure constants. Choosing $a_1=a_2=\o{a}{\sqrt{2}}\ne 0$ we have
the usual curvature definitions for a de Sitter algebra with
cosmological constant $\Lambda= \varepsilon a^2$,
namely the superextension of the
$SL (2, R)$ Lie algebra. In the limit $a_2 \to 1$ and $a_1 \to 0$ we
get the usual $D=2$ analogue of the
$N=2$ super Poincar\'e Lie algebra, where, calling $L$ the
$U(1)$ generator dual to the graviphoton, the supercharges $Q^\pm ,
\t Q_\pm$ are neutral under $L$,
\begin{equation}
[L,Q^\pm ]= [ L,\tilde Q_\pm ]=0.
\label{2}
\end{equation}
In this case, the generator $L$ can be interpreted
as a ``central charge", since
it appears in the supercharges anticommutators:
\be
\{ Q^- , \t Q_- \} \sim  L, \qquad
\{ Q^+ , \t Q_+ \} \sim L. \ee
Finally, in the limit $a_2 \to 0$, $a_1 \to 1$ we get a new kind of
Poincar\'e superalgebra, named by us ``charged Poincar\'e ", where the
supercharges do rotate under the $U(1)$ action:
\begin{equation}
[ L,Q^\pm ]=\pm Q^\pm,\quad\quad [ L,\tilde Q_\pm ]=\mp \tilde Q_\pm,
\end{equation}
In this case $L$ is not a central charge, since it does not appear in the
supercharge anticommutators. Indeed, one has
\ba
\{ Q^+ , Q^- \} = P, \quad &\quad & \{ \tilde Q_+ , \tilde Q_- \}=
\tilde P, \nn\\
\{ Q^+ , \t Q_+ \} = 0, \quad &\quad & \{ Q^- , \tilde Q_- \}=0,
\ea
$P$ and $\t P$ being the left and right translations, dual to $e^+$ and
$ e^-$ respectively.

In \cite{billofre}
the construction of global $N=2$ supersymmetric theories was
based on the use of the ordinary Poincar\'e superalgebra. In this case
we can always choose the gauge $\omega =A=0$ and we can altogether
forget about these one forms. In the solution of Bianchi identities we
simply have to respect global Lorentz and U(1) symmetries. However
the flat case is actually unable to distinguish between the ordinary
and charged Poincar\'e algebra. At the level of curved superspace,
on the other hand,
there is a novelty that distinguishes $D=2$ from
higher dimensions. It turns out that the correct algebra is the
charged one.

The field content of
the off shell graviton multiplet is easily described. The zweibein
describes one bosonic degree of freedom (four components restricted to
one by two diffeomorphisms and by the Lorentz symmetry), while each
gravitino describes two degrees of freedom
(four components restricted by two supersymmetries). Finally, the
graviphoton  $A$ yields one bosonic degree of freedom (two components
restricted by the $U(1)$ gauge symmetry). The mismatch of two bosonic degrees
of freedom is filled by a complex scalar auxiliary field $M$ and by
its conjugate $\bar M$. The problem is therefore that of writing a
rheonomic parametrization for the curvatures (\ref{rhbi}) using as free
parameters  their space-time components plus an auxiliary complex
scalar $M$.

As can be easily read from (\ref{rhbi}) the curvature two forms satisfy:
\ba
\nabla T^+ &=& Re^+ - \o{i}{2} (\rho^+ \zeta^- -\zeta^+ \rho^-),\nn\\
\nabla T^- &=& -Re^- - \o{i}{2} \varepsilon
(\tilde\rho_+ \tilde\zeta_- -\tilde\zeta_+ \tilde\rho_-),
\nn\\
\nabla \rho^\pm &=& \o{1}{2} R \zeta^\pm \pm\o{ia_1}{4} F \zeta^\pm
+a_1 a_2 (\tilde\rho^\mp e^+ -\tilde\zeta_\mp T^+ ),\nn\\
\nabla \tilde\rho_\pm &=&- \o{1}{2} R \tilde\zeta_\pm \mp\o{ia_1}{4}
F \tilde\zeta_\pm
-a_1 a_2 \varepsilon (\rho^\mp e^- - \zeta^\mp T^-),\nn\\
\nabla R &=& -2 a_1^2 a_2^2\varepsilon
(T^+e^- -e^+T^- )-\o{i}{2}a_1 a_2
(\rho^+ \tilde\zeta_+ -\zeta^+ \tilde\rho_+ + \rho^- \tilde\zeta_- -\zeta^-\t
\rho^- ),\nn\\
\nabla F &=& -a_2
( \rho^- \tilde\zeta_- -\zeta^-\tilde\rho_- -\rho^+ \tilde\zeta_+ +\zeta^+
\tilde\rho_+).
\label{bianchigrav}
\ea

The general solution for the above Bianchi identities with vanishing
torsions is
\ba
\begin{array}{ll}
T^+ = 0, &\quad\quad
T^- = 0, \cr
\rho^+ = \tau^+ e^+ e^- -a_1 (M-a_2) \tilde\zeta_- e^+, & \quad\quad
\tilde\rho_+= \tilde\tau_+ e^+ e^- +a_1  \varepsilon (M -a_2)\zeta^- e^-,\cr
\end{array}\nonumber\\
\begin{array}{l}
R = ({\cal R} -2a_1^2a_2^2\varepsilon )e^+ e^- +
\o{i}{2}\varepsilon  e^-(\tau^+ \zeta^- + \tau^- \zeta^+)
+ \o{i}{2} e^+ (\tilde\tau_-  \tilde\zeta_+ +\tilde\tau_+ \t
\zeta^-) \cr
\phantom{R=}+\o{ia_1}{2}
\left [(M-a_2) \zeta^- \tilde\zeta_- + (\bar M -a_2)\zeta^+ \t
\zeta^+ \right ],\cr
F ={ \cal F} e^+ e^- + (M-a_2) \zeta^- \tilde\zeta_- - (\bar M -a_2)\zeta^+
\tilde\zeta_+ -\o{1}{a_1} \varepsilon (\tau^+ \zeta^- - \tau^- \zeta^+)e^- \cr
\phantom{R=}+\o{1}{a_1}(\tilde\tau_-
\tilde\zeta_+ -\tilde\tau_+ \tilde\zeta_-)e^+.
\end{array}
\label{rheograv}
\ea
The formulae for $\rho^-$ and $\tilde\rho_-$
can be derived from those of $\rho^+$ and $\tilde\rho_+$ by
complex conjugation. In doing this, one has to keep into account that
the complex conjugation reverses the order of the fields in a product
of fermions.

It is immediate to see in eq.s. (\ref{rheograv})
that the limit $a_1 \to 0$ is singular, and this
reflects the fact that we are not able to find the correct
parametrizations for this case.
On the contrary the limit $a_2 \to 0$ is perfectly
consistent and we call it "charged Poincar\'e algebra".

{}From now on, to avoid any confusion in using the formulae
for the curvature
definition, {\it we will always refer to the symbols
$R,F, \rho^\pm,\tilde\rho^\pm$
as to the ones defined in (\ref{rhbi}) with }$a_1=1, a_2=0$.

The general rule for obtaining the solution to the Bianchi identities is the
rheonomic principle. We briefly describe it in three steps.

i) One expands the curvatures two-forms $\rho^\pm$,
$\tilde\rho^\pm$, $R$ and $F$ in a basis of superspace two-forms:
the ``spacetime'' form $e^+e^-$ and the ``superspace'' forms, which
can be fermionic, like
$\zeta^\pm e^\pm$ and $\tilde\zeta^\pm e^\pm$,
or bosonic, like $\zeta^\pm\zeta^\pm$,
$\zeta^\pm\tilde\zeta^\pm$ and $\tilde\zeta^\pm\tilde\zeta^\pm$.

ii) The coefficients $\tau^\pm$, $\tilde\tau^\pm$, ${\cal R}$ and ${\cal
F}$ of the spacetime form $e^+e^-$ are independent ones: the rheonomic
parametrizations (\ref{rhbi}) can be viewed as a definition of them.
They are the supercovariantized derivatives of
the fields. In particular, ${\cal R}$ is the supercurvature
and ${\cal F}$ is the super-field-strength.

iii) The coefficients of the superspace forms, instead, are functions of
the fields and of the supercovariantized derivatives
$\tau^\pm$, $\tilde\tau^\pm$, ${\cal R}$ and ${\cal F}$.
They are determined by solving the Bianchi identities (\ref{bianchigrav})
in superspace. Their form is strongly constrained by Lorentz
invariance, global $U(1)$
invariance and scale invariance.
These restrictions are such that the role of (\ref{bianchigrav}) is
simply that of fixing some numerical coefficients, while providing
also several self-consistency checks.
Moreover, imposition of (\ref{bianchigrav}) also provides the
rheonomic parametrizations of
$\nabla \tau^\pm$, $\nabla \tilde\tau^\pm$, $\nabla {\cal R}$ and
$\nabla{\cal F}$ and
of the covariant derivatives $\nabla M$
and $\nabla \bar M$ of the auxiliary fields $M$ and $\bar M$, namely
\begin{eqnarray}
\nabla\tau^+&=&\nabla_+\tau^+e^++\nabla_-\tau^+e^-+
\left({1\over 2}{\cal R}+{i\over 4}{\cal F} -\varepsilon M\bar M\right)\zeta^+
+\nabla_-M\tilde\zeta_-,\nonumber\\
\nabla\tilde\tau_+&=&\nabla_+\tilde\tau_+e^++\nabla_-\tilde\tau_+e^--
\left({1\over 2}{\cal R}+{i\over 4}{\cal F} -\varepsilon
M\bar M\right)\tilde\zeta_+
+\nabla_+M\zeta_-,\nonumber\\
\nabla M&=&\nabla_+ Me^++\nabla_- Me^--{i\over 2}\varepsilon
(\tilde\tau_+\zeta^++\tau^+\tilde\zeta_+),\nonumber\\
\nabla{\cal R}&=&\nabla_+{\cal R}e^++\nabla_-{\cal R}e^-+
{i\over 2}(\varepsilon \nabla_-\tilde\tau_-\tilde\zeta_++\varepsilon
\nabla_-\tilde\tau_+
\tilde\zeta_- -\nabla_+\tau^-\zeta^+ -\nabla_+\tau^+\zeta^-)\nonumber\\
&-& i [\bar M (\tau^+ \tilde\zeta_+ + \tilde\tau_+
\zeta^+)+M(\tau^-\tilde\zeta_- + \tilde\tau_- \zeta^- )],\nonumber\\
\nabla{\cal F}&=&\nabla_+{\cal F}e^++\varepsilon \nabla_-{\cal F}e^- -
\varepsilon \nabla_-\tilde\tau_-\tilde\zeta_+ +\nabla_-\tilde\tau_+
\tilde\zeta_-+\nabla_+\tau^-\zeta^+-\nabla_+\tau^+\zeta^-.
\label{residuo}
\end{eqnarray}
These equations, in their turn, are the definitions of the
supercovariantized derivatives of $\tau^\pm$, $\tilde\tau^\pm$, ${\cal
R}$, ${\cal F}$, $M$ and $\bar M$.
Finally, the $e^+e^-$ sector of the Bianchi
identities (\ref{bianchigrav}) gives the ``space-time'' counterparts
of the Bianchi identities themselves, i.e.\ the formul\ae\ for
$[\nabla_+,\nabla_-]\Phi$ of any field $\Phi$.

The formal correspondence between de Sitter and anti de Sitter
theories is summarized by
\begin{equation}
e^-\rightarrow -e^-,\quad \tau\rightarrow -\tau,\quad {\cal R}\rightarrow
-{\cal R},\quad {\cal F}\rightarrow -{\cal F},\quad \nabla_-\rightarrow
-\nabla_-.
\end{equation}
{}From (\ref{residuo}) we can confirm that $\varepsilon =1$
corresponds to positive curvature, while $\varepsilon =-1$
corresponds to negative
curvature. Indeed, setting ${\cal R}$=const and ${\cal F}=0$,
the expressions of $\nabla {\cal R}$ and $\nabla {\cal F}$
imply either $M=\bar M=0$
or $\tau^\pm=\tilde\tau^\pm=0$. If $M=\bar M=0$, then $\nabla M$
and $\nabla \bar M$
also imply $\tau^\pm=\tilde\tau^\pm=0$. So, we can conclude that
$\tau^\pm=\tilde\tau^\pm=0$ is in any case true. Finally,
$\nabla \tau$ implies ${\cal R} =2\varepsilon M\bar M$
and $M=$const. This also shows that one cannot move from the de Sitter
to the anti de Sitter case by a continuous deformation of the expectation
value of $M,\bar M$.

For simplicity, from now on we set $\varepsilon=+1$.

\section{Coupling gravity with chiral matter}
\label{coupling}

In the previous section we have derived the first ingredient we need,
namely the off-shell graviton multiplet structure. In the present
section we extend the rheonomic construction of chiral multiplets
\cite{LG} discussed in \cite{billofre} for flat superspace,
to the curved superspace environment.
The field content of an off-shell chiral multiplet is $X^I, \psi^I ,
\tilde\psi^I, H^I $ where $X^I$ is a complex scalar field,
$\psi^I ,\tilde\psi^I$ are complex spin $\pm\o{1}{2}$ fields
and $H^I$ is a complex auxiliary
scalar field. The complex conjugate fields will be denoted by a star.
The index notation is $I=(0,i)$, $i=1,\cdots n$.
The multiplet corresponding to the value $I=0$ plays a
special role in coupling to supergravity, namely it is the multiplet
containing the Lagrange multiplier $\Phi$ introduced in eq.s
(\ref{intro923}) and (\ref{intro24}).

To start our program we need the
covariant derivatives for the matter fields
\footnotemark\footnotetext{Our notation for the covariant derivative is
$\nabla \phi= d \phi - s\omega \phi -\o{i}{2}q A$, where $s,q$ are
the spin and the $U(1)$ charge for the field $\phi$}, which are
\ba
\nabla X^I &=& d X^I, \nn\\
\nabla \psi^I &=& d\psi^I -\o{1}{2} \omega \psi^I + \o{i}{4} A \psi^I,
\nn\\
\nabla \tilde\psi^I &=& d \tilde\psi^I +
\o{1}{2}\omega \tilde\psi^I -\o{i}{4}A \tilde\psi^I, \nn\\
\nabla H^I &=& dH^I.
\label{cova}
\ea
{}From the Bianchi identities, which are easily read off (\ref{cova}), we find
the following rheonomic pa\-rame\-tri\-za\-tions:
\ba
\nabla X^I &=& \nabla_+ X^I e^+ + \nabla_- X^I e^- +
\psi^I \zeta^- +\tilde\psi^I \tilde\zeta_-, \nn\\
\nabla\psi^I &=& \nabla_+\psi^I e^+ + \nabla_-\psi^I e^- -\o{i}{2}
\nabla_+ X^I \zeta^+ + H^I \tilde\zeta_-,\nn\\
\nabla \tilde\psi^I &=& \nabla_+ \tilde\psi^I e^+ + \nabla_-
\tilde\psi^I e^- -\o{i}{2}
\nabla_- X^I \tilde\zeta_+ - H^I  \zeta^-, \nn\\
\nabla H^I &=& \nabla_+ H^I e^+ + \nabla_- H^I e^- -
\o{i}{2} \nabla_- \psi^I \tilde\zeta_+
+\o{i}{2} \nabla_+ \tilde\psi^I \zeta^+.
\label{rheomatter}
\ea
The usual choice for the auxiliary field in the Landau-Ginzburg matter is
\be
H^I= \eta^{I \js} \partial_\js \bar W,
\label{auxi}
\ee
$\eta_{IJ^*}$ denoting a flat (constant) metric and $W(X)$ being a
(polynomial) chiral potential.
If we explicitly make this choice,
we also find the fermionic equation of
motions from the self-consistency of the parametrization $\nabla H^I$:
\ba
&\o{i}{2}& \nabla_- \psi^I - \eta^{I \js} \partial_{M^*} \partial_{J^*}
\bar W \t
\psi^{M^*}=0, \nonumber\\
&\o{i}{2}&\nabla_+ \tilde\psi^I +\eta^{I \js}\partial_{M^*}\partial_{j^*}
\bar W
\psi^{M^*}=0.
\label{auxi2}
\ea
Finally, from the supersymmetric variations of the fermionic field equation
we find the bosonic field equation
\ba
[ \nabla_- \nabla_+ &+& \nabla_+\nabla_- ]X^I - 8\eta^{I \js} \partial_{M^*}
\partial_{\js} \partial_{L^*} \bar W \psi^{L^*} \tilde\psi^{M^*} +8
\eta^{I \js}\partial_{M^*} \partial_{\js} \bar W \eta^{M^* L}\partial_L W \nn\\
&-&4i \bar M \eta^{I \js}\partial_\js \bar W
+ \tau^- \psi^I - \tilde\tau_- \tilde\psi^I=0.
\label{auxi3}
\ea
The coupling of the Landau-Ginzburg matter with N=2 supergravity
is described
by the following Lagrangian, derived from the field equations (\ref{auxi}),
(\ref{auxi2}) and (\ref{auxi3})
\be
{\cal L}_{Liouville}= {\cal L}_{kin} + {\cal L}_{W},
\ee
where ${\cal L}_{kin}$ and ${\cal L}_W$ are
the kinetic and superpotential terms
\ba
{\cal L}_{kin} &=& \eta_{I\js} (\nabla X^I - \psi^I
\zeta^- - \tilde\psi^I \tilde\zeta_-
) (\Pi^{\js}_+ e^+ - \Pi_-^\js e^- )\nn\\
&+& \eta_{I\js} (\nabla X^{\js} + \psi^{\js}
\zeta^+ + \tilde\psi^\js \tilde\zeta_+) (\Pi^{I}_+ e^+ - \Pi_-^I e^- )\nn
\\
&+& \eta_{I\js}  (\Pi^{I}_+ \Pi_-^{\js}+  \Pi_-^I\Pi_+^\js )e^+ e^-
\nn\\
&+&2i  \eta_{I\js} (-\psi^I \nabla \psi^\js e^+ - \psi^\js \nabla \psi^I e^+
+ \tilde\psi^I \nabla \tilde\psi^\js e^- + \tilde\psi^\js \nabla
\tilde\psi^I e^- )\nn\\
&+& \eta_{I\js}(\nabla X^\js \psi^I \zeta^- - \nabla X^I \psi^\js \zeta^+
- \nabla X^\js \tilde\psi^I \tilde\zeta_- + \nabla X^I \tilde\psi^\js
\tilde\zeta_+)\nn\\
&+& \eta_{I\js} (\psi^I \tilde\psi^\js \zeta^- \tilde\zeta_+ + \psi^\js
\tilde\psi^I \tilde\zeta_- \zeta^+ ) -8 \eta_{I \js}H^I H^\js e^+ e^-,
\nonumber\\
{\cal L}_W&=& 4i (\psi^I \partial_I W \tilde\zeta_+ e^+
+\psi^\js \partial_{\js} \bar W \tilde\zeta_- e^+
+\tilde\psi^I \partial_I W \zeta^+ e^- + \tilde\psi^\is \partial_\is
\bar W \zeta^- e^-)\nn\\
&+& 8 [ (\partial_I \partial_J W \psi^I \tilde\psi^J - \partial_{\is}
\partial_\js \bar W
\psi^\is\tilde\psi^\js ]e^+e^-\nn\\
&+& 4i(MW-\bar M\bar W)e^+ e^- +2 \bar W \tilde\zeta_- \zeta^- - 2 W
\tilde\zeta_+
\zeta^+ \nn\\
&+& (8 H^I \partial_I W + 8H^\is \partial_\is \bar W) e^+ e^-.
\label{matterlagr}
\ea
The fields $\Pi_\pm^I$ and $\Pi_\pm^{I*}$ are auxiliary fields for
the first order formalism: their equation of motion equates them to
the supercovariant derivatives of the $X$-fields,
\begin{equation}
\Pi_\pm^I=\nabla_\pm X^I, \hskip 2truecm
\Pi_\pm^{I*}=\nabla_\pm X^{I*}.
\end{equation}
Substitution of these expressions in ${\cal L}_{kin}$ gives the usual
second order Lagrangian. The rheonomic parametrizations
of $\nabla\Pi_\pm^I$ and $\nabla\Pi_\pm^{I*}$ are derived from the Bianchi
identities and the rheonomic parametrizations (\ref{rheomatter}),
in the same way as (\ref{residuo}) are derived from the Bianchi identities
(\ref{bianchigrav}) and the rheonomic parametrizations (\ref{rheograv}).

\section{N=2 Liouville gravity}
\label{liouville}

Let us consider the chiral multiplet labelled with the index $I=0$.
We call it ``dilaton" multiplet.
For convenience, we relabel the dilaton multiplet as
\be
(X^0,X^{0^*}, \psi^0, \psi^{0^*},\tilde\psi^0,\tilde\psi^{0^*}, H^0, H^{0^*})
\to (X, \bar X, \lambda_-, \lambda_+, \tilde\lambda^-,
\tilde\lambda^+ , H, \bar H ). \nn
\ee
The $N=2$ extension of the Lagrangian $(X +\bar X)R$ is given by
\begin{eqnarray}
{\cal L}_1 &=& (X + \bar X)R - \o{i}{2}(X - \bar X) F
- 2 \lamb_- \rho^- + 2 \lamb_+ \rho^+
+ 2 \tilde\lamb^- \tilde\rho_- - 2\tilde\lamb^+
\tilde\rho_+ \nonumber\\&&
- 4i \bar M H e^+ e^- + 4i M \bar H e^+ e^-.
\label{lagra}
\end{eqnarray}
Let us remind the reader how  a supersymmetric Lagrangian is
constructed in the rheonomic
framework. It is sufficient to find an ${\cal L}$
that satisfies
\begin{equation}
\nabla {\cal L}=d{\cal L}=0.
\label{rheocond}
\end{equation}
In checking this equation one has to use
the rheonomic parametrizations
(\ref{rheograv}) and (\ref{rheomatter}) together with the
definitions (\ref{rhbi}) and (\ref{cova}).
${\cal L}_1$ was determined
starting from the first term $(X+\bar X)R$ and guessing the other
ones in order to satisfy (\ref{rheocond}).

One can pass from the second order formalism to the first order one
by adding the term
\begin{equation}
{\cal L}_T=p_+ T^+ +p_- T^- ,
\end{equation}
where $p_+, p_-$ are (bosonic) Lagrangian multipliers implementing the
torsion constraint $T^\pm=0$. ${\cal L}_T$ is clearly
supersymmetric (the supersymmetry variation of the spin connection
is still determined from the variations of zweibein and gravitini:
this is the so-called {\sl 1.5 order formalism}).
Moreover, one can add to eq. (\ref{lagra}) a ``cosmological constant
term"
compatible with the N=2 local supersymmetry
\begin{eqnarray}
{\cal L}_2 &=& (MX + \bar M\bar X) e^+ e^-
+\lamb_-  \tilde\zeta_+ e^+ - \lamb_+ \tilde\zeta_- e^+ +  \tilde\lamb^-
\zeta^+ e^-  - \tilde
\lamb^+
\zeta^- e^- \nn\\
&+& {i\over 2} X
\zeta^+ \tilde\zeta_+ + {i\over 2}\bar X\zeta^- \tilde\zeta_-
+ 2i (\bar H -H)e^+ e^-,
\label{desitter}
\ea
so that the total Lagrangian is ${\cal L}={\cal L}_1 + {\cal L}_2$.

The equations for the auxiliary fields are
\begin{equation}
H= -\o{i}{4} \bar X,  \quad\quad  \bar H =\o{i}{4}X, \quad\quad
M=\bar M = -\o{1}{2}.
\end{equation}
Using the equation of motions of
$H$, $\bar H$ and $X+\bar X$,
we get precisely a de Sitter supergravity with
cosmological constant $\Lambda=\o{1}{2}$. The field strength $F$, on
the other hand, is set to zero by the $X-\bar X$ field equation.

Notice that when in the matter Lagrangian (\ref{matterlagr})
the index $I$ takes the value $0$ and $W=-\o{i}{4} X^0$,
then ${\cal L}_W$ coincides with the de Sitter term ${\cal L}_2$.

To conclude this section, let us show that the kinetic term of the
dilaton multiplet can be produced from the Poincar\'e Lagrangian
${\cal L}_1$ with some field redefinitions.
One can perform the substitutions
\begin{eqnarray}
\begin{array}{ll}
e^+\rightarrow e^+ {\rm e}^{-{1\over 4}(X+\bar X)},\quad & \quad
e^-\rightarrow e^- {\rm e}^{-{1\over 4}(X+\bar X)},\cr
M\rightarrow (M+iH){\rm e}^{{1\over 2}(X+\bar X)},\quad & \quad
\bar M\rightarrow (\bar M-i\bar H){\rm e}^{{1\over 2}(X+\bar X)},\cr
\zeta^+\rightarrow {\rm e}^{-{1\over 4}X}(\zeta^+-i\lambda_- e^+),\quad & \quad
\zeta^-\rightarrow {\rm e}^{-{1\over 4}\bar X}(\zeta^-+i\lambda_+ e^+),\cr
\tilde\zeta_+\rightarrow {\rm e}^{-{1\over 4}X}(\tilde\zeta_+-i\tilde
\lambda^- e^-),\quad & \quad
\tilde\zeta_-\rightarrow {\rm e}^{-{1\over 4}\bar X}(\tilde\zeta_-+i
\tilde\lambda^+ e^-),\cr
\lambda_-\rightarrow \lambda_-{\rm e}^{{1\over 4}\bar X},\quad & \quad
\lambda_+\rightarrow \lambda_+{\rm e}^{{1\over 4}X},\cr
\tilde\lambda^-\rightarrow \tilde\lambda^-{\rm e}^{{1\over 4}\bar X},
\quad & \quad
\tilde\lambda^+\rightarrow \tilde\lambda^+{\rm e}^{{1\over 4}X},\cr
\end{array}
\end{eqnarray}
and
\begin{eqnarray}
\Omega &\rightarrow & \Omega-{1\over 2}[
\nabla_+\bar Xe^+-\nabla_-\bar Xe^--\lambda_+\zeta^+
+\tilde\lambda^+\tilde\zeta_+],\nonumber\\
\bar \Omega &\rightarrow & \bar \Omega-{1\over 2}[
\nabla_+Xe^+-\nabla_-Xe^-+\lambda_-\zeta^-
-\tilde\lambda^-\tilde\zeta_-],
\label{2222}
\end{eqnarray}
where $\Omega=\omega-{i\over 2}A$, $\bar \Omega=
\omega+{i\over 2}A$.
Then, the Poincar\'e Lagrangian ${\cal L}_1$ (\ref{lagra}) goes into
\begin{equation}
{\cal L}_1+{\cal L}_{kin},
\end{equation}
${\cal L}_{kin}$ being the kinetic Lagrangian of the dilaton multiplet,
the first formula of
(\ref{matterlagr}) with $I$ restricted to the value $0$.
As one can easily check,
the replacement of $\omega={1\over 2}(\Omega+\bar \Omega)$
implied by (\ref{2222})
is consistent with the preservation of vanishing torsions.

\section{Gauge free algebra of the N=2 Liouville theory}
\label{gaugefree}

The local symmetries of the N=2 Liouville theory are:
diffeomorphism, local lorentz rotations,
supersymmetries and U(1) gauge symmetry.
The procedure for writing down the
free BRST algebra \cite{ansfre,ansfre2,bonora}
is straightforward, once the curvature
definitions and the rheonomic parametrizations are BRST extended
to ghost forms, by introducing the ghosts of the local symmetries:
\ba
\begin{array}{ll}
\hat e^+ = e^+ + C^+, \quad & \quad
\hat e^- = e^- + C^-, \\
\hat \zeta^\pm = \zeta^\pm + \Gamma^\pm, \quad & \quad
\hat {\tilde{\zeta}}_\pm = \tilde\zeta_\pm + \tilde\Gamma_\pm,  \\
\hat \omega = \omega +C^0, \quad & \quad
\hat A =A + C. \\
\end{array}
\ea
The exterior derivative is BRST extended to $\hat d= d+ s$.
The BRST variations of the fields are easily derived  by
selecting out the correct ghost number sector in the BRST extensions
of formul\ae\ (\ref{rhbi}), (\ref{rheograv})
(graviton multiplet) and (\ref{cova}), (\ref{rheomatter})
(dilaton multiplet).
We have, in particular, for the graviton multiplet,
\ba
s e^+ &=& -\nabla C^+ - C^0 e^+ +\o{i}{2} \zeta^+ \Gamma^-
+\o{i}{2} \Gamma^+ \zeta^-, \nonumber\\
s e^- &=& -\nabla C^- + C^0 e^- +\o{i}{2} \tilde\zeta_+ \tilde\Gamma_-
+\o{i}{2} \tilde\Gamma_+ \tilde\zeta_-,\nonumber\\
s \zeta^+ &=&-\nabla \Gamma^+ - \o{1}{2} C^0 \zeta^+ -\o{i}{4}C
\zeta^+ + \tau^+ (C^+ e^- + e^+ C^-)
- M (\tilde\Gamma_- e^+ + \tilde\zeta_- C^+),  \nonumber\\
s \tilde\zeta_+ &=&-\nabla \tilde\Gamma_+ + \o{1}{2}
C^0 \tilde\zeta_+ +\o{i}{4}
C \tilde\zeta_++ \tilde\tau_+( C^+ e^- +  e^+ C^-)
+ M (\Gamma^- e^- + \zeta^- C^-),  \nonumber\\
s\omega &=&-dC^0+
{\cal R}(C^+ e^-+ e^+ C^-)+
\o{i}{2}C^-(\tau^+\zeta^-+\tau^-\zeta^+)+
\o{i}{2}C^+(\tilde\tau_+\tilde\zeta_-+\tilde\tau_-\tilde\zeta_+)
\nonumber\\&&+\o{i}{2}e^-(\tau^+\Gamma^-+\tau^-\Gamma^+)
+\o{i}{2}e^+(\tilde\tau_+\tilde\Gamma_-+\tilde\tau_-\tilde\Gamma_+)
\nonumber\\&&
+\o{i}{2}M(\zeta^-\tilde\Gamma_-+\Gamma^-\tilde\zeta_-)+
\o{i}{2}\bar M(\Gamma^+
\tilde\zeta_++\zeta^+\tilde\Gamma_+),\nonumber\\
sA &=&-dC+{\cal F}(C^+ e^-+ e^+ C^-)-(\tau^+\zeta^--\tau^-\zeta^+)C^-
\nonumber\\&&
+(\tilde\tau_-\tilde\zeta_+-\tilde\tau_+\tilde\zeta_-)C^+
-(\tau^+\Gamma^--\tau^-\Gamma^+)e^- \nonumber\\&&+
(\tilde\tau_-\tilde\Gamma_+-\tilde\tau_+\tilde\Gamma_-)e^+
\nonumber\\&&+M(\zeta^-\tilde\Gamma_-+\Gamma^-\tilde\zeta_-)-\bar M(\Gamma^+
\tilde\zeta_+ +\zeta^+\tilde\Gamma_+).
\ea

On the other hand,
the BRST transformations of the fields of the dilaton multiplet are
\begin{eqnarray}
sX&=&\nabla_+X C^++\nabla_-X C^-+\lambda_-\Gamma^-+
\tilde\lambda^-\tilde\Gamma_-,\nonumber\\
s\lambda_-&=&\nabla_+\lambda_-C^++\nabla_-\lambda_-C^-
-{i\over 2}\nabla_+X\Gamma^++H\tilde\Gamma_-,\nonumber\\
s\tilde\lambda^-&=&\nabla_+\tilde\lambda^-C^++\nabla_-\tilde\lambda^-C^-
+{i\over 2}\nabla_+X\tilde \Gamma_+-H\Gamma^-,\nonumber\\
sH&=&\nabla_+HC^++\nabla_-HC^--{i\over 2}
\nabla_-\lambda_-\tilde\Gamma_++{i\over 2}\nabla_+\tilde\lambda^-
\Gamma^+.
\end{eqnarray}

\section{Topological Twist of the N=2 Liouville Theory}
\label{twist}

In this section we perform the topological twist of the theory.
The formal set-up is analogous to the one in four
dimensions \cite{ansfre,ansfre2}. One has to change consistently,
the spin, the BRST charge and the ghost number.
In particular, the new BRST charge is obtained by the so-called
{\sl topological shift}, which is a simple redefinition of the supersymmetry
ghosts that get ghost number zero. Due to the existence of
Majorana-Weyl spinors in two dimensions, one has two possibilities,
known in the literature as the A and B twists \cite{index1,billofre}.

The geometrical and physical
meaning of the two types of twists was discovered at the
level of globally supersymmetric N=2 matter theories.
As noticed in eq.\ (\ref{matterlagr}), the most general
interaction of a set of chiral multiplets involves
two separately supersymmetric Lagrangian terms,
the kinetic term
${\cal L}_{kin}$ and the superpotential term
${\cal L}_{W}$. The choice between the A and B twists decides
which term is BRST nontrivial and which one
is BRST exact. In the A twist the nontrivial BRST cohomology is
carried by ${\cal L}_{kin}$,
while in the B twist it is carried by ${\cal L}_W$.
In the first case, the topologically meaningful coupling parameters are those
corresponding to the K\"ahler class deformations of the target space metric,
the correlation functions being instead independent of the
deformation parameters of the superpotential. In the B twist
the situation is reversed.

The above considerations apply both to globally
and locally supersymmetric theories. However, in presence of supergravity
and as far as the A twist is concerned,
${\cal L}_W$  cannot be set to zero from the beginning,
since it contains the
de Sitter term ${\cal L}_2$ (\ref{desitter}) (formally
obtainable from ${\cal L}_W$ with $W=-{i\over 4}X^0$) and
the parameter $a$ that can be put
in front of ${\cal L}_2$ is the cosmological constant,
the sign of which has to
be compatible with the Euler characteristic $\int R=2(1-g)$.

Technically, the A and B twists emerge as follows.
We first
notice that the Lagrangian (\ref{lagra}) of Poincar\'e  gravity,
possesses a global $R$-symmetry
[which will be denoted by $U(1)^\prime$], under which
the fields transform with the charges shown in table \ref{topotable}.
$U(1)^\prime$ is not a local symmetry and it is not even a global
symmetry for the de Sitter Lagrangian (\ref{desitter}).
In general R-symmetries and R-dualities play a crucial role
\cite{ansfre2} in the
topological twist of the N=2 theories.
Depending on the choice of the twist (A or B), the new Lorentz group is
defined as a combination of the old one with the $U(1)^\prime$
or $U(1)$ symmetry; viceversa for the ghost number.

For the A twist the new assignments  and the topological shift are
\begin{eqnarray}
\begin{array}{ll}
{\rm spin}^\prime = \hbox{spin} + U(1)^\prime, \quad &\quad
\Gamma^+ \to  \Gamma^+ + \alpha,\nonumber\\
{\rm ghost}^\prime = {\rm ghost} + 2 U(1),\quad &\quad
\tilde\Gamma_- \to  \tilde\Gamma_- +\beta,
\end{array}
\label{spina}
\end{eqnarray}
while for the B twist they are
\begin{eqnarray}
\begin{array}{ll}
{\rm spin}^\prime = \hbox{spin} + U(1), \quad &\quad
\Gamma^+ \to  \Gamma^+ + \alpha, \nonumber\\
{\rm ghost}^\prime = {\rm ghost} + 2 U (1)^\prime,
\quad &\quad
\tilde\Gamma_+ \to  \tilde\Gamma_+ + \beta.
\end{array}
\label{spinb}
\end{eqnarray}
$\alpha$ and $\beta$ are the so-called {\sl brokers} \cite{ansfre2}.
They are to be treated formally as constant ($d\alpha=d\beta=0$) and
their (purely formal) role is to bring the correct contributions
of spin and ghost number to the fields (see the last column of
table \ref{topotable}). Their quantum numbers are given in the table.

In this paper we focus on the A twist.
The topological theory that emerges from our
analysis is apt to perform the gravitational dressing of the topological
theories dealing  with K\"ahler class deformations \cite{billofre}.
The gravitational dressing of the complex structure deformations need the B
twist of the N=2 Liouville theory, whose analysis is postponed to future work.

The shift produces a new BRST operator $s^\prime$ which
equals $s+\delta_T$, $\delta_T$ being the topological variation
(known as ${\cal Q}_s$ in conformal field theory). On the
graviton multiplet, $\delta_T$ acts as
\begin{eqnarray}
\begin{array}{ll}
\delta_T e^+ ={i\over 2}\alpha \zeta^-, \quad & \quad
\delta_T e^- ={i\over 2} \tilde \zeta_+\beta, \\
\delta_T\zeta^+ = - {1\over 2} \omega \alpha - {i\over 4} A \alpha - M
\beta e^+ \equiv B_1\alpha,
\quad & \quad
\delta_T \zeta^- =0,\\
\delta_T \tilde\zeta_+=0,
\quad & \quad
\delta_T \tilde\zeta_- =  {1\over 2} \omega \beta - {i\over 4}
A \beta + \bar M \alpha e^- \equiv B_2\beta,\\
\delta_T M=-{i\over 2}\tilde\tau_+\alpha ,\quad &\quad
\delta_T \bar M=-{i\over 2}\tau^-\beta,
\end{array}\nonumber\\
\begin{array}{l}
\delta_T\omega = {i\over 2}M\zeta^-\beta +
{i\over 2}\bar M\alpha \tilde\zeta_+ +{i\over 2}e^-\tau^-\alpha
+{i\over 2}e^+\tilde\tau_+\beta,
\quad\quad\quad\quad\quad\quad\quad\quad\\
\delta_T A = M\zeta^-\beta -\bar M\alpha
\tilde\zeta_+ +\tau^-\alpha e^--\tilde\tau_+\beta e^+.
\quad\quad\quad\quad\quad\quad\quad\quad
\end{array}
\label{deltaT}
\end{eqnarray}
Taking into account that the BRST algebra closes
off-shell, we see that $B_1$
and $B_2$ play the role of Lagrange multipliers, since they are
the BRST variations of the antighosts $\zeta^+$ and $\tilde\zeta_-$.
$B_1$ and $B_2$ can be considered as redefinitions of $A$, $M$ and
$\bar M$. Indeed, since $M$ and $\bar M$ have spin $1$ and $-1$ after the
twist,
$Me^+$ and $\bar M e^-$ can be considered as one forms. In particular,
we have shown that the graviphoton $A$ belongs to
${\cal B}_{gauge-fixing}$. On the other hand, it is clear that
the gauge-free topological algebra is that of $SL(2,{\bf R})$,
since the above formul\ae\ show that the topological symmetry
is the most continuous
deformation of the zweibein.

On the dilaton multiplet, $\delta_T$ is
\begin{eqnarray}
\begin{array}{ll}
\delta_T X = \tilde\lambda^- \beta, \quad & \quad
\delta_T {\bar X} =  - \lambda_+ \alpha, \\
\delta_T \lambda_- = -{i\over 2} \nabla_+ X \alpha + H \beta
\equiv H_1\alpha,\quad & \quad
\delta_T \tilde\lambda^- = 0, \\
\delta_T \lambda_+ = 0, \quad & \quad\delta_T \tilde\lambda^+ =
{i\over 2} \nabla_- {\bar X} \beta - \bar H \alpha\equiv H_2\beta, \\
\delta_T H ={i\over 2} \nabla_+ \tilde\lambda^- \alpha, \quad & \quad
\delta_T {\bar H} = - {i\over 2} \nabla_- \lambda_+ \beta.\\
\end{array}
\label{deltaT2}
\end{eqnarray}
$H_1$ and $H_2$ are also Lagrange multipliers, redefinitions
of $H$ and $\bar H$.

Finally, the topological variation of the brokers vanishes, but
nilpotence of $s^\prime$ and $s$ requires
\begin{eqnarray}
\begin{array}{ll}
s^\prime \alpha =- {1\over 2}
C^0 \alpha - {i\over 4} C \alpha=s\alpha, \quad & \quad
s^\prime \beta =  {1\over 2} C^0 \beta - {i\over 4} C \beta=s\beta.
\end{array}
\end{eqnarray}
In other words, even if formally, $\alpha$ and $\beta$ have to be considered
as sections with definite spin and $U(1)$ charge.

{}From the above formul\ae\ it is simple to check that $\delta_T$ is
nilpotent, $\delta_T^2=0$, as expected. Summarizing, we have
\begin{equation}
s^\prime=\delta_T+s,\quad\quad s^{\prime \, 2}=s^2=\delta_T^2=s\delta_T+
\delta_T s=0.
\end{equation}

Using the above formulae and
the notation shown in the last column of
table \ref{topotable}, we can write the
full Lagrangian $\cal L$ as the topological variation of a suitable
gauge fermion $\Psi$ plus a total derivative term. Precisely,
\begin{equation}
{\cal L}_1=\delta_T\Psi_1+2\nabla(XM_-e^-+\bar X M_+e^+),
\,\quad\,
{\cal L}_2=\delta_T\Psi_2+\nabla(\bar Xe^+-X e^-),
\end{equation}
where
\begin{eqnarray}
\Psi_1&=&-2X d\bar{\tilde \xi}+2\bar X d\bar \xi-4i\chi_+M_-e^+e^-
-4i\chi_-M_+e^+e^-+2XB_2\bar{\tilde \xi}-2\bar X B_1\bar \xi,\nonumber\\
\Psi_2&=&(\bar X e^+-Xe^-)(\bar{\tilde \xi}-\bar \xi)-2i(\chi_++\chi_-)
e^+e^-.
\end{eqnarray}
{}From the formula of $\Psi_2$, it is apparent that $U(1)^\prime$
and correspondingly spin$^\prime$
are violated by ${\cal L}_2$. On the other hand, ${\cal L}_2$
is purely a gauge-fixing term, so that this violation does
not affect the Lorentz
symmetry in the physical correlators. It can be thought
as the choice of a noncovariant gauge-fixing.

Let us analyze the gauge-fixing conditions of the twisted theory.
We take ${\cal L}_{tot}={\cal L}_1-2a{\cal L}_2$, ${\cal L}_1$
and ${\cal L}_2$ being given by (\ref{desitter}).
The Lagrange multipliers $A$, $M$-$\bar M$ and $H$-$\bar H$
impose the following constraints
\begin{equation}
\nabla_+(X-\bar X)=\nabla_-(X-\bar X)=0,\quad \,\,
H={i\over 2}a\bar X,\quad \,\, \bar H=-{i\over 2} a X,\quad \,\,
M=\bar M=a.
\label{constr}
\end{equation}
A check of consistency is that the $\delta_T$ variations of the constraints
(\ref{constr}) imposed by the Lagrange multipliers are the field
equations of the topological ghosts $\zeta^-$, $\tilde\zeta_+$,
$\lambda_+$ and $\tilde\lambda^-$, obtained from the variations
of ${\cal L}_{tot}$
with respect to the corresponding antighosts
$\lambda_-$, $\tilde\lambda^+$, $\zeta^+$ and $\tilde\zeta_-$, i.e.\
\begin{eqnarray}
\begin{array}{ll}
\tau^-=0,\quad &\quad
\tilde \tau_+=0,\cr
\nabla_+\lambda_-=0,\quad &\quad
\nabla_-\lambda_-=a\tilde\lambda^+,\cr
\nabla_+\tilde\lambda^-=-a\lambda_+,\quad &\quad
\nabla_-\tilde\lambda^-=0.
\end{array}
\end{eqnarray}
To verify this, one has to keep into account that the
$\omega$ field equation gives
$p_+=-\nabla_+(X+\bar X)$, and $p_-=\nabla_-(X+\bar X)$.

The observables of the topological theory are easily derived, as in
the case of the Verlinde and Verlinde model, from the descent
equations $\hat d \hat R^n=0$, $\hat R=R+\psi_0+\gamma_0$ being
the BRST extension of the curvature $R$. In particular, the local
observables are
\begin{equation}
\sigma^{(0)}_n(x)=\gamma_0^n(x),
\end{equation}
as anticipated in the introduction. On the other hand, the field
strength $F$ does not provide any new observables, due to the fact
that $A\in {\cal B}_{gauge-fixing}$.

\section{The conformal field theory associated with N=2 Liouville gravity}
\label{conformal}

In this section, we analyse N=2 Liouville gravity in detail. We gauge-fix it
and show that, in the limit $a^2\rightarrow 0$, it reduces to a
conformal field theory of vanishing total central charge,
summarized by formul\ae\ (\ref{6.30}) and (\ref{ghrepr}).
The total
central charge is the sum of the central charge of the Liouville system,
equal to $6$, and that of the ghost system, equal to $-6$.

We start from the rheonomic Lagrangian of Poincar\'e N=2 D=2 supergravity,
that we rewrite here for convenience,
\begin{equation}
{\cal L}_1=(X+\bar X) R-{i\over 2}(X-\bar X)F-2
\lambda_- \rho^-+
2 \lambda_+\rho^++2\tilde\lambda^- {\tilde\rho}_--2
\tilde\lambda^+
{\tilde\rho}_++4i(M\bar H-\bar M H)E^zE^{\bar z}.
\end{equation}
In this section, we use the notation $E^z$ and $E^{\bar z}$ instead
$e^+$ and $e^-$.
We do this for the sake of a tensorial notation that is useful in
the gauge fixed theory. We shall have tensor indices $t^{z\cdots z{\bar
z}\cdots{\bar z}}$ that are raised and lowered with the flat metric
${\hat g}_{z{\bar z}} ={\hat g}^{z{\bar z}}=1$ and spinor indices
$s^\pm$ and $\tilde s_\pm$ such that $s^{+-} \sim t^z$ and
$\tilde s_{+-} \sim t^{\bar z}$. In this way it is immediate to read
the spin assignments of the fields.

We also recall that in order to deal with the torsions, we have to
add a term
\begin{equation}
\pi_zT^z+\pi_{\bar z}T^{\bar z},
\label{tors}
\end{equation}
that allows to treat the spin connection $\omega$ as an independent
variable.

In the case of locally supersymmetric theories, one has to perform
the topological twist
on the BRST quantized version of the theory,
as discussed in detail in ref.\ \cite{ansfre}.
We have developed the full gauge-free BRST algebra
of N=2 Liouville theory in section \ref{gaugefree}. Now we proceed to gauge-fix
diffeomorphisms, Lorentz rotations, supersymmetries and local $U(1)$
gauge transformations. Then we discuss the gauge-fixed BRST theory.

Diffeomorphisms and Lorentz rotations are fixed by choosing the conformal gauge
\begin{equation}
\matrix{E^z\wedge dz=0,&E^z\wedge d\bar z+E^{\bar z}\wedge
dz=0,&E^{\bar z}\wedge d\bar z=0.}
\label{diffgf}
\end{equation}
These conditions permit to express the zweibein as
\begin{equation}
\matrix{E^z={\rm e}^{\varphi(z,\bar z)} dz,&E^{\bar z}=
{\rm e}^{\varphi(z,\bar z)}d\bar z,}
\label{gfcond1}
\end{equation}
where $\varphi(z,\bar z)$ is the conformal factor, which is to be
identified with the Liouville quantum field.

Supersymmetries are fixed by extending the conformal gauge by means of
the conditions
\begin{equation}
\matrix{\zeta^+\wedge E^z=0,\,\, &\quad
\zeta^-\wedge E^z=0,\,\, &\quad\tilde\zeta_+\wedge E^{\bar z}=0,
\,\, &\quad\tilde\zeta_-\wedge E^{\bar z}=0,}
\label{susygf}
\end{equation}
that, together with (\ref{diffgf}) make the so-called
superconformal gauge. (\ref{susygf}) corresponds to the usual condition
$\gamma^\mu\zeta^A_\mu=0$, $A=1,2$ and permit to
express the gravitini as
\begin{equation}
\matrix{\zeta^+=\eta^+_z {\rm e}^\varphi dz,&
\zeta^-=\eta^-_z  {\rm e}^\varphi dz,&
\tilde\zeta_+=\eta_{+\bar z} {\rm e}^\varphi d\bar z,&
\tilde\zeta_-=\eta_{-\bar z} {\rm e}^\varphi d\bar z,}
\label{gfcond2}
\end{equation}
where $\eta^\pm_z(z,\bar z)$ and $\eta_{\bar z}^\pm(z,\bar z)$
are anticommuting fields of spin $1/2$ and $-1/2$ (the superpartners
of the Liouville field $\varphi$).

The $U(1)$ gauge transformations have to be treated carefully.
The critical N=2 string possesses {\sl two} local $U(1)$
gauge-symmetries, that permit to gauge-fix the graviphoton $A$ to zero
and introduce the two $b$-$c$ systems {\sl a la} Faddeev-Popov,
one in the left moving sector and one in the right
moving one.
In such a way, a complete chiral factorization
into two superconformal field theories (left and right moving) is achieved.
The theory that we are now dealing with, on the other hand,
possesses a single local $U(1)$ symmetry, the $U(1)^\prime$ R-symmetry
being only global. From a field theoretical point of view,
it is not immediate to see how a pair of $b$-$c$ systems can
be introduced {\sl a la} Faddeev-Popov. Indeed, enforcing the
usual Lorentz gauge $\partial^\mu A_\mu=0$ produces
a second order ghost-antighost system. That is why we pay
a particular attention to this fact. We arrange things in the correct way
by using a trick.
Let us introduce an additional trivial BRST system
(a ``one dimensional topological $\sigma$-model'')
$\{\xi,C^\prime\}$, $\xi$ being a ghost number zero scalar
and $C^\prime$ being a ghost number one scalar.
Their BRST algebra is chosen to be
trivial, namely
\begin{equation}
\matrix{s\xi=C^\prime,&sC^\prime=0.}
\label{trivial}
\end{equation}
The meaning of this BRST system is the gauging
of the R-symmetry $U(1)^\prime$.
Indeed, $U(1)^\prime$, which is only a {\sl global} symmetry of the
starting theory, becomes a {\sl local} symmetry
in the gauge-fixed version of the same theory.
This will become clear later on, when the complete factorization
between left and right moving sectors will be apparent.
We fix both the $U(1)$ gauge symmetry and the trivial
symmetry (\ref{trivial}) by choosing the following two gauge-fixings
\begin{equation}
\matrix{A_z-\partial_{z}\xi=0,&A_{\bar z}+\partial_{\bar z}\xi=0.}
\label{1.7}
\end{equation}
corresponding to $A=*d\xi$, where $A=A_zdz+A_{\bar z}d\bar z$.

With obvious notation, the $\bar C$ and $\bar \Gamma$ fields
being antighosts, the gauge-fermion $\Psi$ is
\begin{eqnarray}
\Psi&=&\Psi_{diff}+\Psi_{susy}+\Psi_{gauge}\nonumber\\&&=
\bar C_{zz}E^z\wedge dz +\bar C_{z\bar z}(E^z\wedge d\bar z+E^{\bar z}\wedge
dz)+\bar C_{\bar z\bar z}E^{\bar z}\wedge d\bar z \nonumber\\&&+
\bar\Gamma_{+ z}\zeta^+\wedge E^z+\bar\Gamma_{-z}\zeta^-\wedge E^z
+\bar\Gamma^+_{\bar z}\tilde\zeta_+\wedge E^{\bar z}+
\bar\Gamma^-_{\bar z}\tilde\zeta_-\wedge E^{\bar z}\nonumber\\&&+
\bar C\wedge (A-*d\xi),
\end{eqnarray}
$\bar C$ being a one form, $\bar C=\bar C_z dz + \bar C_{\bar z}d\bar z$.

The gauge-fixed Poincar\'e Lagrangian is thus
\begin{equation}
{\cal L}={\cal L}_1+s\Psi.
\end{equation}

In writing the gauge-fixed Lagrangian we proceed as follows.
The Lagrange multipliers of the algebraic gauge-fixings (i.e.\
diffeomorphisms, Lorentz rotations and supersymmetries) will
be functionally integrated away, thus solving the gauge-fixing conditions
(\ref{diffgf}) and (\ref{susygf}). The
Lagrange multipliers $P_z$ and $P_{\bar z}$ of the $U(1)$ and
$U(1)^\prime$ gauge-fixings ($s\bar C_z=P_z$ and $s\bar C_{\bar z}=P_{\bar z}$)
will be conveniently retained for now,
since the corresponding gauge-fixings (\ref{1.7}) contain
derivatives of the fields.
The remaining part of the Lagrangian can thus be
greatly simplified by using the algebraic gauge-fixing conditions.
Moreover, noticing that $\zeta^+\zeta^-=\tilde\zeta_+\tilde\zeta_-=0$
on the gauge-fixing condition (\ref{gfcond2}), the torsion constraints
imposed by the Lagrange multipliers $\pi_z$ and $\pi_{\bar z}$
become algebraic ``gauge-fixing conditions'' on $\omega$
that are solved by
\begin{equation}
\omega=\partial_{z} \varphi \, dz-\partial_{\bar z} \varphi \, d\bar z.
\end{equation}
Notice that such an $\omega$ does not depend on the gravitini.
The Lorentz ghost $C^0$
appears only algebraically. Consequently,
we can eliminate $\bar C_{z\bar z}$, by expressing $C^0$ in terms of the other
fields
\begin{eqnarray}
2{\rm e}^\varphi C^0 dz\wedge d\bar z &=&-\nabla C^z  \wedge d\bar z
-\nabla C^{\bar z}   \wedge dz\nonumber\\&&+{i\over 2}
[(\Gamma^{+}\zeta^-+\zeta^+\Gamma^{-}) \wedge d\bar z
+(\tilde\zeta_+\tilde\Gamma_{-}+\tilde\Gamma_{+}\tilde\zeta_-)
 \wedge dz].
\label{varepsilon}
\end{eqnarray}
Due to this, one finds
\begin{eqnarray}
s\Psi_{diff}+s\Psi_{susy}&=&\bar C_{zz}\nabla C^z   \wedge dz+
\bar C_{\bar z\bar z}\nabla C^{\bar z}  \wedge d\bar z\nonumber\\&&+
\bar\Gamma_{+z}\nabla (\Gamma^{+} E^z+\zeta^+ C^z )
+\bar\Gamma_{-z}\nabla(\Gamma^{-} E^z+\zeta^- C^z )
\nonumber\\
&&+\bar \Gamma^+_{\bar z}\nabla(\tilde\Gamma_{+} E^{\bar z}+
\tilde \zeta_+ C^{\bar z} )
+\bar\Gamma^-_{\bar z}\nabla(\tilde\Gamma_{-} E^{\bar z}+
\tilde\zeta_- C^{\bar z} ).
\end{eqnarray}

It is convenient to introduce the following substitutions
\begin{eqnarray}
\begin{array}{ll}
\eta^{\pm\prime}_z=\eta^\pm_z{\rm e}^{{1\over 2}\left(\varphi\mp
{i\over 2}\xi\right)},&\quad
\eta^\prime_{\pm\bar z}=\eta_{\pm\bar z}
{\rm e}^{{1\over 2}\left(\varphi\mp{i\over 2}\xi\right)},\\
\lambda^\prime_\pm=\lambda_\pm{\rm e}^{{1\over 2}
\left(\varphi\pm{i\over 2}\xi\right)},&\quad
\tilde\lambda^{\pm\prime}=\tilde\lambda^\pm{\rm e}^{{1\over 2}
\left(\varphi\pm{i\over 2}\xi\right)},\\
C^{z\prime}= C^z  {\rm e}^{-\varphi},&\quad
C^{\bar z\prime}= C^{\bar z}  {\rm e}^{-\varphi},\\
\bar C_{zz}^\prime=\bar C_{zz} {\rm e}^{\varphi},&\quad
\bar C_{\bar z\bar z}^\prime=\bar C_{\bar z\bar z} {\rm e}^{\varphi},\\
\Gamma^{\pm\prime}=\Gamma^\pm
{\rm e}^{-{1\over 2}\left(\varphi\pm{i\over 2}\xi\right)}
-\eta^{\pm\prime}_zC^{z\prime},&\quad
\tilde\Gamma^\prime_\pm=
\tilde\Gamma_\pm\,
{\rm e}^{-{1\over 2}\left(\varphi\pm{i\over 2}\xi\right)}
-\eta^\prime_{\pm\bar z}C^{\bar z\prime},\\
\bar\Gamma_{\pm z}^\prime=\bar\Gamma_{\pm z}{\rm e}^
{{3\over 2}\varphi\pm{i\over 4}\xi},&\quad
\bar\Gamma_{\bar z}^{\pm\prime}=\bar\Gamma_{\bar z}^\pm{\rm e}^
{{3\over 2}\varphi\pm{i\over 4}\xi},\\
\end{array}
\end{eqnarray}
which are also allowed in the functional integral, since the
Jacobian determinant is one.

The gauge-fixed versions of the gravitini curvatures give
\begin{equation}
\tau^++M\eta_{-\bar z}=-{\rm e}^{-{3\over 2}\varphi+
{i\over 4}\xi}
\partial_{\bar z}\eta^{+\prime}_z,
\end{equation}
and similar relations, that provide gauge-fixed
formul\ae\ for the supercovariantized derivatives.
On the other hand, the field equation of $\omega$ gives expressions
for $\pi_z$ and $\pi_{\bar z}$, that will be useful for computing the BRST
charge ${\cal Q}_{BRST}$. An alternative way of finding $\pi_z$
and $\pi_{\bar z}$ is that of imposing the independence of ${\cal Q}_{BRST}$
from $C^0$.

With $\pi=1/2 \, (X+\bar X)$ and
$\chi=i/2 \, (X-\bar X)$, we have
\begin{eqnarray}
{\cal L}_1+s\Psi_{diff}+s\Psi_{susy}&=&
-4\pi\partial_{z} \partial_{\bar z}\varphi
+\chi(\partial_{\bar z} A_z-\partial_{z} A_{\bar z})
-2\lambda^{\prime}_+\partial_{\bar z}\eta^{+\prime}_z
+2\lambda^{\prime}_-\partial_{\bar z}\eta^{-\prime}_z\nonumber\\&&
-2\tilde\lambda^{+\prime}\partial_{z} \eta^{\prime}_{+\bar z}
+2\tilde\lambda^{-\prime}\partial_{z}\eta^{\prime}_{-\bar z}
+\bar C_{zz}^\prime\partial_{\bar z} C ^{z\prime}
-\bar C_{\bar z\bar z}^\prime\partial_{z} C^{\bar z\prime}\nonumber\\&&+
\bar\Gamma_{+z}^\prime\partial_{\bar z}\Gamma^{+\prime}
+\bar\Gamma_{-z}^\prime\partial_{\bar z}\Gamma^{-\prime}
-\bar\Gamma_{\bar z}^{+\prime}\partial_{z} \tilde\Gamma_+^\prime
-\bar\Gamma_{\bar z}^{-\prime}\partial_{z} \tilde\Gamma_-^\prime
\nonumber\\
&+&{i\over 4}(A_z-\partial_{z} \xi)(2\tilde\lambda^{+\prime}
\eta^{\prime}_{+\bar z}+2\tilde\lambda^{-\prime}\eta^{\prime}_{-\bar z}
+\bar\Gamma^{+\prime}_{\bar z}\tilde\Gamma_+^\prime-
\bar\Gamma_{\bar z}^{-\prime}\tilde\Gamma_-^\prime)
\nonumber\\&-&{i\over 4}(A_{\bar z}+\partial_{\bar z}\xi)
(2\lambda^{\prime}_+
\eta^{+\prime}_z+2\lambda^{\prime}_-\eta^{-\prime}_z-
\bar\Gamma_{+z}^\prime\Gamma^{+\prime}+
\bar\Gamma_{-z}^\prime\Gamma^{-\prime}).
\end{eqnarray}

Now, it remains to deal with the term $s\Psi_{gauge}$. We find
\begin{eqnarray}
s\Psi_{gauge}&=&P_z(A_{\bar z}-\partial_{\bar z}\xi)-
P_{\bar z}(A_z-\partial_{z} \xi)\nonumber\\&&
-\bar C_z[\partial_{\bar z}(C+C^\prime)+C^{z\prime}
(\partial_{z} A_{\bar z} -\partial_{\bar z} A_z)
+\partial_{\bar z}\eta^{+\prime}_z\Gamma^{-\prime}-\partial_{\bar z}
\eta^{-\prime}_z
\Gamma^{+\prime}]\nonumber\\&&
+\bar C_{\bar z}[\partial_{z} (C-C^\prime)-C^{\bar z\prime}(
\partial_{z} A_{\bar z}-\partial_{\bar z} A_z)
+\partial_{z}\eta^{\prime}_{-\bar z}\tilde\Gamma_+^\prime
-\partial_{z}
\eta^{\prime}_{+\bar z}\tilde  \Gamma^{\prime}_-]
\nonumber\\&&+{i\over 4}\bar C_{\bar z}(A_z-\partial_{z}\xi)(
\eta^{\prime}_{-\bar z}\tilde\Gamma_+^\prime
+\eta^{\prime}_{+\bar z}\tilde\Gamma_-^\prime)
\nonumber\\&&-{i\over 4}\bar C_z
(A_{\bar z}+\partial_{\bar z}\xi)(\eta^{+\prime}_z\Gamma^{-\prime}
+\eta^{-\prime}_z\Gamma^{+\prime}).
\end{eqnarray}
Integrating away both $P_z$, $P_{\bar z}$ and $A_z$, $A_{\bar z}$,
we can use the corresponding field equations.
Defining
\begin{eqnarray}
\begin{array}{ll}
c=C+C^\prime-2C^{z\prime}\partial_{z} \xi+\eta^{+\prime}_z
\Gamma^{-\prime}-\eta^{-\prime}_z\Gamma^{+\prime},&\quad\quad
\bar C_z=b_z,\\
\bar c=C-C^\prime+2C^{\bar z\prime}\partial_{\bar z} \xi-
\eta^{\prime}_{+\bar z}
\tilde\Gamma_-^\prime+\eta^{\prime}_{-\bar z}\tilde\Gamma_+^\prime,&
\quad\quad
\bar C_{\bar z}=b_{\bar z},
\end{array}\nonumber\\
\begin{array}{ll}
\bar C_{zz}^{\prime\prime}=
\bar C_{zz}^\prime-2b_z\partial_{z} \xi,&\quad
\bar C_{\bar z\bar z}^{\prime\prime}=
\bar C_{\bar z\bar z}^\prime+2b_{\bar z}\partial_{\bar z}\xi,\quad\quad\quad\\
\bar\Gamma_{+z}^{\prime\prime}=\bar\Gamma_{+z}^\prime-b_z
\eta^{-\prime}_z,&\quad
\bar\Gamma_{-z}^{\prime\prime}=\bar\Gamma_{-z}^\prime+b_z
\eta^{+\prime}_z,\quad\quad\quad\\
\bar\Gamma_{\bar z}^{+\prime\prime}=\bar\Gamma_{\bar z}^{+\prime}+
b_{\bar z}\eta^{\prime}_{-\bar z},&\quad
\bar\Gamma_{\bar z}^{-\prime\prime}=\bar\Gamma_{\bar z}^{-\prime}-
b_{\bar z}\eta^{\prime}_{+\bar z},\quad\quad\quad\\
\end{array}
\label{s2}
\end{eqnarray}
the total gauge-fixed Poincar\'e Lagrangian takes the form
\begin{eqnarray}
{\cal L}_{Poincar\grave e}&=&-4\pi\partial_{z} \partial_{\bar z}\varphi+
2\chi\partial_{z} \partial_{\bar z}\xi
-2\lambda^{\prime}_+\partial_{\bar z}\eta^{+\prime}_z
+2\lambda^{\prime}_-\partial_{\bar z}\eta^{-\prime}_z\nonumber\\&&
-2\tilde\lambda^{+\prime}\partial_{z} \eta^{\prime}_{+\bar z}
+2\tilde\lambda^{-\prime}\partial_{z} \eta^{\prime}_{-\bar z}
+\bar C_{zz}^{\prime\prime}\partial_{\bar z} C^{z\prime}
-\bar C_{\bar z\bar z}^{\prime\prime}\partial_{z} C^{\bar z\prime}
-b_z\partial_{\bar z} c+b_{\bar z} \partial_{z}\bar c
\nonumber\\&&
+\bar\Gamma_{+z}^{\prime\prime}\partial_{\bar z}\Gamma^{+\prime}
+\bar\Gamma_{-z}^{\prime\prime}\partial_{\bar z}\Gamma^{-\prime}
-\bar\Gamma_{\bar z}^{+\prime\prime}\partial_{z} \tilde\Gamma_+^\prime
-\bar\Gamma_{\bar z}^{-\prime\prime}\partial_{z} \tilde\Gamma_-^\prime.
\label{lapoi}
\end{eqnarray}
Again, the Jacobian determinant
corresponding to (\ref{s2}) is one.

It is natural to conjecture that
Poincar\'e N=2 supergravity corresponds to an N=2 superconformal
field theory. To show that this is indeed the case, we compute
the energy-momentum tensor $T_{zz}$, the supercurrents
$G_{+z}$ and $G_{-z}$, the $U(1)$ current $J_z$.
In order to do this, we first compute the BRST charge
${\cal Q}^{BRST}=\oint J_z^{BRST}dz$,
$J^{BRST}_z$ denoting the left moving BRST current.
Acting with ${\cal Q}^{BRST}$ on the various antighost fields
it is then simple to derive the ``gauge-fixings'', which are, in our
case, the N=2 currents. Since the BRST symmetry
is a global symmetry and the gauge-fixed action is BRST-invariant,
the BRST current $J^{BRST}$ can be found by performing a {\sl local}
BRST transformation $\delta_{BRST}=\kappa(x)\, s$, where $s$ is
the BRST operator and $\kappa (x)$ is a point-dependent ghost number
$-1$ scalar parameter. The variation of ${\cal L}$ can
then be expressed as
\begin{equation}
\delta_{BRST}{\cal L}=*d\kappa\wedge J^{BRST}=
(\partial_z\kappa \,\,J_{\bar z}^{BRST}+\partial_{\bar z}\kappa \,\,
J_z^{BRST})\, dz
\wedge d \bar z.
\end{equation}
As anticipated, expressions for $\pi_z$ and $\pi_{\bar z}$ can be found
by requiring the independence of $J^{BRST}$ from $C^0$. In particular,
one finds
\begin{eqnarray}
\pi_z{\rm e}^\varphi &=&-2\partial_z\pi+
\lambda_-^\prime\eta_z^{-\prime}
-\lambda_+^\prime\eta_z^{+\prime}+\bar C_{zz}^\prime C^{z\prime}
-{1\over 2}b_z(\eta_z^{+\prime}\Gamma^{-\prime}-\eta_z^{-\prime}
\Gamma^{+\prime})
\nonumber\\&&
+{1\over 2}(\bar \Gamma_{+z}^\prime\Gamma^{+\prime}+
\bar \Gamma_{-z}^\prime\Gamma^{-\prime})-
(\bar \Gamma_{+z}^\prime\eta^{+\prime}_z+
\bar \Gamma_{-z}^\prime\eta^{-\prime}_z)C^{z\prime}.
\end{eqnarray}
We separate the Liouville and ghost sectors by writing
\begin{equation}
\matrix{T_{zz}=T_{zz}^{grav}+T_{zz}^{gh},&J_z=J_{z}^{grav}+J_{z}^{gh},\cr
G_{+z}=G_{+z}^{grav}+G_{+z}^{gh},&G_{-z}=G_{-z}^{grav}+G_{-z}^{gh},}
\end{equation}
and similarly for the complex conjugates $T_{\bar z\bar z}$, $J_{\bar z}$,
$\bar G_{\bar z}^+$ and $\bar G_{\bar z}^-$.
Let us first focus on the Liouville sector. We have, on shell
and up to total derivative terms,
\begin{eqnarray}
J_z^{BRST\, grav}&=&-C^{z\prime}T_{zz}^{grav}-{i\over 4}
cJ_{z}^{grav}+
{i\over 2}\Gamma^{+\prime}G_{+z}^{grav}+
{1\over 2}\Gamma^{-\prime}G_{-z}^{grav},
\end{eqnarray}
where, after a simple redefinition of the fields
\begin{eqnarray}
\begin{array}{llll}
\varphi\rightarrow \varphi,\quad&\quad \pi\rightarrow {1\over 4}
\pi,\quad&\quad\xi\rightarrow -2i\xi,\quad&\quad\chi\rightarrow
{i\over 4}\chi,\cr
\lambda^{\prime}_+\rightarrow -{i\over 4}\lambda_+,\quad&\quad
\lambda_-^\prime\rightarrow {1\over 4}\lambda_-,\quad&\quad
\eta^{+\prime}_z\rightarrow 2i\eta^+_z,\quad&\quad
\eta^{-\prime}_z\rightarrow 2 \eta^-_z,
\end{array}
\end{eqnarray}
the N=2 currents are written as
\begin{eqnarray}
T_{zz}^{grav}&=&-\partial_{z} \pi\partial_{z}\varphi+
{1\over 2}\partial_{z}^2\pi+\partial_{z}
\chi\partial_{z}\xi+{1\over 2}(\partial_{z}\lambda_-\eta^-_z-
\lambda_-\partial_{z}\eta^-_z)+
{1\over 2}(\lambda_+\partial_{z}\eta^+_z-\partial_{z}\lambda_+
\eta^+_z),\nonumber\\
G_{+z}^{grav}&=&\partial_{z}\lambda_+-\lambda_+\partial_{z}(\varphi+\xi)
+\eta^-_z\partial_{z}(\chi+\pi),\nonumber\\
G_{-z}^{grav}&=&\partial_{z}\lambda_--\lambda_-\partial_{z}(\varphi-\xi)
+\eta^+_z\partial_{z}(\chi-\pi),\nonumber\\
J_{z}^{grav}&=&\partial_{z}\chi-\lambda_-\eta^-_z-\lambda_+\eta^+_z.
\label{6.30}
\end{eqnarray}
The background charge term $1/2 \,\, \partial_{z}^2\pi$ has
no influence on the central charge, that turns out to be $c_{grav}=6$.
This implies that there is also an N=4 conformal symmetry, according
to the analysis of \cite{ALE}.
The fundamental operator product expansions are normalized as follows:
\begin{eqnarray}
\begin{array}{ll}
\partial_{z}\pi(z)\partial_{w}\varphi(w)=-{1\over (z-w)^2},&\quad
\partial_{z}\chi(z)\partial_{w}\xi(w)={1\over (z-w)^2},\\
\lambda_+(z)\eta^+_w(w)=-{1\over z-w},&\quad
\lambda_-(z)\eta^-_w(w)={1\over z-w}.
\end{array}
\end{eqnarray}
It is easy to check that the N=2 operator product
expansions are indeed satisfied by
(\ref{6.30}).

Before going on, let us dwell for a moment on
the above N=2 $c=6$ superconformal algebra and discuss
some of its features. First of all, notice that it
can be decomposed into the
direct sum of two $N=2$ superconformal algebr\ae\ with
central charges $c_1=3/2$ and $c_2=9/2$. We have
\begin{eqnarray}
\matrix{T_{zz}^{grav}=T_{zz}^{(1)}+T_{zz}^{(2)},&
J_{z}^{grav}=J_z^{(1)}+J_z^{(2)},\cr
G_{+z}^{grav}=G_{+z}^{(1)}+G_{+z}^{(2)},&
G_{-z}^{grav}=G_{-z}^{(1)}+G_{-z}^{(2)},}
\end{eqnarray}
where
\begin{eqnarray}
T_{zz}^{(1)}&=&\partial_{z} \varphi_1\partial_{z}
\varphi_1^\prime+{1\over 4}\partial_{z}^2(\varphi_1+
\varphi_1^\prime)+
{1\over 2}(\partial_{z}\lambda_+^{(1)}\lambda_-^{(1)}-\lambda_+^{(1)}
\partial_{z}\lambda_-^{(1)}),\nonumber\\
G_{+z}^{(1)}&=&{1\over \sqrt{2}}
(\partial_{z}\lambda_+^{(1)}+2\lambda_+^{(1)}\partial_{z}\varphi_1),
\quad\quad\quad
G_{-z}^{(1)}={1\over \sqrt{2}}
(\partial_{z}\lambda_-^{(1)}+2\lambda_-^{(1)}\partial_{z}
\varphi_1^\prime),\nonumber\\
J_z^{(1)}&=&{1\over 2}\partial_{z}(\varphi_1-\varphi_1^\prime)
+\lambda_+^{(1)}\lambda_-^{(1)},
\end{eqnarray}
and
\begin{eqnarray}
T_{zz}^{(2)}&=&-\partial_{z} \varphi_2\partial_{z}
\varphi_2^\prime-{1\over 4}\partial_{z}^2
(\varphi_2+
\varphi_2^\prime)-
{1\over 2}(\partial_{z}\lambda_+^{(2)}\lambda_-^{(2)}-\lambda_+^{(2)}
\partial_{z}\lambda_-^{(2)}),\nonumber\\
G_{+z}^{(2)}&=&{1\over \sqrt{2}}
(\partial_{z}\lambda_+^{(2)}+2\lambda_+^{(2)}\partial_{z}\varphi_2),
\quad\quad\quad G_{-z}^{(2)}={1\over \sqrt{2}}
(\partial_{z}\lambda_-^{(2)}+2\lambda_-^{(2)}\partial_{z}\varphi_2^\prime),
\nonumber\\
J_z^{(2)}&=&-{1\over 2}\partial_{z}(\varphi_2-\varphi_2^\prime)
-\lambda_+^{(2)}\lambda_-^{(2)}.
\end{eqnarray}
The correspondence with the previous fields is
\begin{equation}
\matrix{
\varphi_1=-{1\over 2}(\varphi+\xi-\chi-\pi),&
\varphi_1^\prime=-{1\over 2}(\varphi-\xi+\chi-\pi),\cr
\lambda_-^{(1)}={1\over \sqrt{2}}(\lambda_--\eta^+_z),&
\lambda_+^{(1)}={1\over \sqrt{2}}(\lambda_++\eta^-_z),\cr
\varphi_2=-{1\over 2}(\varphi+\xi+\chi+\pi),&
\varphi_2^\prime=-{1\over 2}(\varphi-\xi-\chi+\pi),\cr
\lambda_-^{(2)}={1\over \sqrt{2}}(\lambda_-+\eta^+_z),&
\lambda_+^{(2)}={1\over \sqrt{2}}(\lambda_+-\eta^-_z),}
\end{equation}

Let us recall that unitary representations of the N=2 algebra with $c<
3$ are given by the minimal model series, where $c={3k\over k+2}$
($k\in {\bf N}$), so that our representation $T_{zz}^{(1)}$, $G_{+z}^{(1)}$,
$G_{-z}^{(1)}$,
$J_z^{(1)}$ corresponds to a free field realization of the $k=2$ minimal
model. Construction of these models \cite{N2GKO} as GKO \cite{GKO}
cosets of $SU(2)$ level $k$
supersymmetric Ka\v c-Moody algebra are well known. We also recall
that unitary representations of the N=2 algebra with $c>3$ have
also been obtained as GKO cosets of the $SL(2,{\bf R})$ supersymmetric
Ka\v c-Moody algebra of level $k$, yielding a series $c={3k\over k-2}$
\cite{bars}.
We nickname these representations ``maximal models''.
Therefore, our $T_{zz}^{(2)}$, $G_{+z}^{(2)}$, $G_{-z}^{(2)}$,
$J_z^{(2)}$ representation is a free field realization of the $k=6$
``maximal model''.

Moreover, it is also easy to check that the N=2 $c=6$
superconformal algebra (\ref{6.30}) can be decomposed into the
direct sum of two N=2 $c=3$ superconformal algebr\ae. They correspond
to the subsets of fields $\{(\varphi+\xi)/\sqrt{2},(\chi-\pi)
/\sqrt{2},\lambda_+,\eta^+_z\}$ and
$\{(\varphi-\xi)/\sqrt{2},(\chi+\pi)
/\sqrt{2},\lambda_-,\eta^-_z\}$.

Now, let us come to the ghost currents. We proceed as before, namely,
we first determine the BRST current $J_z^{BRST\, gh}$ from
a local BRST variation of the action and then act with
${\cal Q}^{gh}_{BRST}=\oint J_z^{BRST\, gh}$ on
the antighost fields. One finds
\begin{eqnarray}
J_z^{BRST\, gh}&=&-\partial_{z} C^{z\prime}
\bar C_{zz}^{\prime\prime}C^{z\prime}-{1\over 2}
\partial_{z} C^{z\prime}(\bar\Gamma_{+z}^{\prime\prime}\Gamma^{+\prime}
+\bar\Gamma_{-z}^{\prime\prime}\Gamma^{-\prime})+
C^{z\prime}(\bar\Gamma_{+z}^{\prime\prime}\partial_{z}\Gamma^{+\prime}+
\bar\Gamma_{-z}^{\prime\prime}\partial_{z}\Gamma^{-\prime})
\nonumber\\&&-
{i\over 2}\bar C_{zz}^{\prime\prime}\Gamma^{+\prime}\Gamma^{-\prime}
-{i\over 4}c
(\bar\Gamma_{+z}^{\prime\prime}
\Gamma^{+\prime}-\bar\Gamma_{-z}^{\prime\prime}
\Gamma^{-\prime})+b_zC^{z\prime}\partial_{z} c
\nonumber\\&&
+b_z(\partial_{z}\Gamma^{+\prime}\Gamma^{-\prime}-\Gamma^{+\prime}
\partial_{z}\Gamma^{-\prime}).
\label{arr}
\end{eqnarray}
After the replacements
\begin{equation}
b_z\rightarrow -1/2 \, i b_z,\quad \quad c\rightarrow 2 ic
\end{equation}
and the redefinitions
\begin{equation}
\matrix{\bar C_{zz}^{\prime\prime}= -b_{zz},&C^{+\prime}
=  c^z ,\cr
\bar\Gamma_{+z}^{\prime\prime}= -i\beta_{+z},&\Gamma^{+\prime}
= -i\gamma^+,\cr
\bar\Gamma_{-z}^{\prime\prime}=\beta_{-z},&\Gamma^{-\prime}
= - \gamma^-,}
\end{equation}
one gets
\begin{eqnarray}
T_{zz}^{gh}&=&2b_{zz}\partial_{z} c^z +\partial_{z} b_{zz} c^z +{3\over 2}
\beta_{+z}\partial_{z}\gamma^++{1\over 2}\partial_{z}\beta_{+z}\gamma^++
{3\over 2}
\beta_{-z}\partial_{z}\gamma^-+{1\over 2}\partial_{z}\beta_{-z}\gamma^-+
b_z\partial_{z} c,\nonumber\\
G_{+z}^{gh}&=&3\beta_{+z}\partial_{z} c^z +2\partial_{z}\beta_{+z} c^z
-\gamma^-b_{zz}- \gamma^-\partial_{z} b_z-2  \partial_{z}\gamma^-b_z
-\beta_{+z} c,
\nonumber\\
G_{-z}^{gh}&=&3\partial_{z} c^z \beta_{-z}+2 c^z \partial_{z}\beta_{-z}
-b_{zz}\gamma^++\partial_{z} b_z \gamma^++2 b_z \partial_{z}
\gamma^++c\beta_{-z},
\nonumber\\
J_{z}^{gh}&=&\beta_{-z}\gamma^--\beta_{+z}\gamma^+-2\partial_{z}(b_z
c^z ).
\label{ghrepr}
\end{eqnarray}
The fundamental operator product expansions are
\begin{eqnarray}
\begin{array}{ll}
b_{zz}(z) c^w (w)=-{1\over z-w},&\quad
b_z(z)c(w)=-{1\over z-w},\\
\beta_{+z}(z)\gamma^+(w)={1\over z-w},&\quad
\beta_{-z}(z)\gamma^-(w)={1\over z-w}.\\
\end{array}
\end{eqnarray}
The ghost contribution to the central charge is $c_{gh}=-6$,
so that $c_{tot}=c_{grav}+c_{gh}=0$, as claimed.

Notice that $\beta$ and $\gamma$ commute among themselves, but
anticommute with $b$ and $c$. This is because they carry an odd ghost number
together with an odd fermion number, while $b$ and $c$ carry
zero fermion number and odd ghost number.
In the usual convention, instead, $\beta$ and $\gamma$
commute with everything. The above
currents satisfy the usual N=2 superconformal algebra with both
conventions. This is because we chose an {\sl ad hoc}
ordering between the fields when writing down the supercurrents,
namely $\beta$-$\gamma$ before $b$-$c$ in $G_{+z}^{gh}$ and {\sl vice
versa} in
$G_{-z}^{gh}$. In all other manipulations the double grading
should be taken into account. Only after the topological twist the
fermion number grading is absent (things are correctly arranged by
the broker \cite{ansfre,ansfre2}).

Notice that the ghost sector is made of two
$b$-$c$-$\beta$-$\gamma$ N=1
systems, namely
a system $b_{zz}$-$c^z$-$\beta_{-z}$-$\gamma^-$ with weight
$\lambda_{\beta_{-z}}=3/2$ and a system
$c$-$b_z$-$\gamma^+$-$\beta_{+z}$
with weight $\lambda_{\gamma^+}=-1/2$. These systems
also possess, as it is well-known, an accidental N=2 symmetry
\cite{gliozzi}.
However the standard representation of
the N=2 $c=6$ superconformal algebra made of these
$b$-$c$-$\beta$-$\gamma$ N=1 systems  does not coincide with
(\ref{ghrepr}).

It is interesting to notice that in the new notation, one has
\begin{equation}
J_z^{BRST\, grav}=-c^zT_{zz}^{grav}+{1\over 2}c J_{z}^{grav}
+{1\over 2}\gamma^+{G}_{+z}^{grav}
-{1\over 2}\gamma^-{G}_{-z}^{grav},
\label{star}
\end{equation}
while (\ref{arr}) gives
\begin{eqnarray}
J_z^{BRST\, gh}&=&\partial_{z} c^zb_{zz}c^z+{1\over 2}\partial_{z} c^z
(\beta_{+z}\gamma^++\beta_{-z}\gamma^-)-c^z
(\beta_{+z}\partial_{z}\gamma^++\beta_{-z}\partial_{z}\gamma^-)
-{1\over 2}b_{zz}\gamma^+\gamma^-\nonumber\\&&
-{1\over 2}c(\beta_{+z}\gamma^+-\beta_{-z}\gamma^-)+b_zc^z\partial_{z} c
+{1\over 2}b_z(\partial_{z}\gamma^+\gamma^--\gamma^+
\partial_{z}\gamma^-)\nonumber\\&=&
{1\over 2}\left(-c^zT_{zz}^{gh}+{1\over 2}c J_{z}^{gh}
+{1\over 2}\gamma^+{G}_{+z}^{gh}
-{1\over 2}\gamma^-{G}_{-z}^{gh}\right),
\end{eqnarray}
a formula that is analogous to (\ref{star}),
with the usual ${1\over 2}$ overall factor.
We have omitted some total derivative terms, that are immaterial as far
as ${\cal Q}_{BRST}$ is concerned. Thus
we can also write
\begin{equation}
J_z^{BRST}=-c^z{\cal T}_{zz}
+{1\over 2}c{\cal J}_{z}
+{1\over 2}\gamma^+{\cal G}_{+z}-{1\over 2}\gamma^-
{\cal G}_{-z},
\end{equation}
where
\begin{equation}
\matrix{
{\cal T}_{zz}=T_{zz}^{grav}+{1\over 2}T_{zz}^{gh},&
{\cal J}_{z}=J_{z}^{grav}+{1\over 2}J_{z}^{gh},\cr
{\cal G}_{+z}={G}_{+z}^{grav}+{1\over 2}{G}_{+z}^{gh},&
{\cal G}_{-z}=G_{-z}^{grav}+{1\over 2}G_{-z}^{gh}.}
\end{equation}

The ghost number charge is
\begin{equation}
{\cal Q}_{gh}=\oint b_{zz}c^z+\beta_{+z}\gamma^++\beta_{-z}\gamma^-+b_zc,
\end{equation}
so that ${\cal Q}_{BRST}=\oint J_z^{BRST}$
has ghost number one:
\begin{equation}
[{\cal Q}_{gh},{\cal Q}_{BRST}]={\cal Q}_{BRST}.
\label{exa}
\end{equation}

In the new notation the Lagrangian (\ref{lapoi}) is written as
\begin{eqnarray}
{\cal L}_{Poincar\grave e}&=&
-\pi\partial_{z}\partial_{\bar z}\varphi+\chi\partial_{z}\partial_{\bar z}\xi+
\lambda_-\partial_{\bar z}\eta^-_z-\lambda_+\partial_{\bar z}\eta^+_z
\nonumber\\&&+\tilde\lambda^-\partial_{z}\eta_{-\bar z}-\tilde
\lambda^+\partial_{z}\eta_{+\bar z}-b_{zz}\partial_{\bar z} c^z
+b_{\bar z\bar z}\partial_{z} c^{\bar z}\nonumber\\&&
-\beta_{+z}\partial_{\bar z}\gamma^+-\beta_{-z}\partial_{\bar z}\gamma^-
+\beta_{+\bar z}\partial_{z}\tilde\gamma_++
\beta_{-\bar z}\partial_{z}\tilde\gamma_-
-b_z\partial_{\bar z} c+b_{\bar z} \partial_{z} \bar c.
\label{poinc}
\end{eqnarray}
Let us make a comment about the addition of the kinetic Lagrangian
for the dilaton supermultiplet (see section \ref{liouville}
for an analogous
comment before gauge-fixing). Formula (\ref{matterlagr})
with the index $I$ restricted
only to the value $0$ and with vanishing superpotential $W$ gives,
after gauge-fixing,
\begin{equation}
{\cal L}_{kin}=-{1\over 2}\partial_z\pi\partial_{\bar z}\pi+{1\over 2}
\partial_z\chi\partial_{\bar z}\chi+\lambda_-\partial_{\bar z}\lambda_+
+\tilde\lambda^-\partial_z\tilde\lambda^+
\end{equation}
(a convenient overall numerical factor has been chosen).
It is immediate to see that the redefinitions
\begin{eqnarray}
\begin{array}{ll}
\varphi\rightarrow \varphi-{1\over 2}\pi,&\quad\xi
\rightarrow\xi-{1\over 2}\chi,\\
\eta^\pm_z\rightarrow\eta_z^\pm\mp{1\over 2}\lambda_\mp,&\quad
\eta_{\pm\bar z}\rightarrow\eta_{\pm\bar z}\mp{1\over 2}\tilde\lambda^{\mp},
\end{array}
\end{eqnarray}
turn ${\cal L}_{Poincar\grave e}$ into
${\cal L}_{Poincar\grave e}+{\cal L}_{kin}$.

This completes the program of studying the N=2 algebra associated with
the gauge-fixed Poincar\'e Lagrangian. Before making the topological
twist, we make some comments on the structure of moduli space, zero
modes
and amplitudes.

The global degrees of freedom of the metric are the $3(g-1)$
moduli $m_i$, that are in one-to-one correspondence with
the $3(g-1)$ zero
modes of the spin 2 antighost $b_{zz}$.
There are $g$ ($={\rm dim}\, H^1$) global degrees of freedom (moduli)
$\nu^j$ of the
graviphoton $A$. These moduli are in one-to-one correspondence with the
$g$ zero modes of the spin 1 antighost $b$.
The total number of moduli is thus $4g-3$.
The supermoduli $\hat m$, $\hat \nu$
can be counted by computing the number of zero modes of the
spin $3/2$ antighosts $\beta_{+z}$ and $\beta_{-z}$, that is $4g-4$, one more
than the number of moduli.
The field
$\xi$ describes the local degrees of freedom of $A$
that survive the $U(1)$ gauge-fixing,
in the same way as
$\varphi$ describes the local degrees of freedom of
the metric that survive the gauge-fixing of diffeomorphisms.

Let us write explicitly the form of the amplitudes of the
N=2 Liouville theory:
\begin{eqnarray}
<{\cal O}_1\cdots {\cal O}_n>&=&\int d\Phi
\int_{{\cal M}_g}\prod_{i=1}^{3g-3}dm_id\bar m_i
\int_{{{\bf C}^g / \Lambda}}\prod_{j=1}^gd\nu_jd\bar\nu_j
\int d\hat m d\hat\nu
\nonumber\\&&
\times
\prod_{i=1}^{3g-3}
<\mu^{iz}_{\phantom{i}\bar z}|b_{zz}><\mu^{i\bar z}_{\phantom{i}z}
|b_{\bar z\bar z}>
\prod_{j=1}^g
<\omega^j_{\bar z}|b_{z}><\omega^j_z|b_{\bar z}>\nonumber\\&&\times
\prod_{k=1}^{2g-2}
\delta(<\zeta^{-k}_{\bar z}|\beta_{-z}>)
\delta(<\tilde \zeta^{-k}_z|\beta_{-\bar z}>)
\delta(<\zeta^{+k}_{\bar z}|\beta_{+z}>)
\delta(<\tilde \zeta^{+k}_z|\beta_{+\bar z}>)
\nonumber\\&&
\times M(\lambda,\eta)
c(z_0)\bar c(\bar z_0)\,\prod_i{\rm e}^{{q_i\over 2}\tilde\pi(z_i)}\,
{\rm e}^{-S(m,\hat m,\nu,\hat\nu)}{\cal O}_1\cdots {\cal O}_n.
\label{n=2amplitudes}
\end{eqnarray}

Let us explain the notation.

i) $\mu^{iz}_{\phantom{i}\bar z}$ denote the $3g-3$ Beltrami differentials,
while $\mu^{i\bar z}_z$ are their complex conjugates.
$<\mu^z_{\bar z}|b_{zz}>=\int_{\Sigma_g}
\mu^z_{\bar z}b_{zz}d^2z$ are the correct insertions that
take care of the $b_{zz}$ zero modes \cite{dhoker} and, at the same time,
take into account of the Jacobian determinant (Beltrami differential)
coming from the change of variables $dg_{\mu\nu}\rightarrow \prod_i
dm_id\bar m_i$.

ii) $\zeta^{\pm k}_{\bar z}$ and $\tilde \zeta^{\pm k}_z$ are
the super Beltrami differentials.
The integration $\int d\hat m d\hat \nu$
over supermoduli (that usually produce the
supercurrent insertions) is not explicitly performed, since $S(m,
\hat m,\nu,\hat\nu)$ does not depend trivially on them and moreover
the observables ${\cal O}_i$ possibly depend on them.

iii) $\omega^j_z$ ($\omega^j_{\bar z}$) are the (anti-)holomorphic
differentials parametrizing
the global degrees of freedom of the graviphoton $A$.
The $g$-dimensional moduli space
of $A$ is the Jacobian variety \cite{kra}
${{\bf C}^g\over \Lambda}$, $\Lambda={\bf Z}^g+\Omega
{\bf Z}^g$ denoting the lattice $\nu_j\approx
\nu_j+n_j+m_k\Omega_{kj}$, $n_j$ and
$m_k$ being integer numbers and $\Omega_{jk}$ being the period matrix.
The reason why one has to restrict the integration of the $U(1)$
moduli to the unit cell is the same as the one that enforces the
restriction of the integration on metrics to the proper moduli space
$T/\Gamma$, $T$ denoting the Teichmuller space and $\Gamma$ the
mapping class group. Indeed, ${\bf C}^g$ is the Teichmuller space
parametrizing the deformations of $A$ that are orthogonal to gauge
transformations. There are, however, gauge
transformations that are not connectible to the identity
and the homotopy classes of these are in one-to-one correspondence
with the Jacobian lattice ${\bf Z}^g+\Omega {\bf Z}^g$.
The $U(1)$ gauge transformations that are
not connectible to the identity correspond to
the shifts $\nu_j\rightarrow\nu_j+n_j+m_k\Omega_{kj}$.
For fixed Chern class, two gauge equivalent
$U(1)$ connections $A_1$ and $A_2$
are such that $A_{1}-A_{2}$ can be
written as $U^{-1}dU$ for $U={\rm e}^{i\phi}$, $\phi:\Sigma_g
\rightarrow S^1$ being a map that winds $n_j$ times around the $B_j$
cycles and $m_k$ times around the $A_k$ cycles.

iv) $M(\lambda,\eta)$ generically denotes the insertions that are necessary
in order to remove the zero modes of the $\eta$'s and the $\lambda$'s
(insertions that can be provided by the observables ${\cal O}_i$),
while $c(z_0)\bar c(\bar z_0)$ is the insertion for the (constant)
zero modes of $c$ and $\bar c$. It is understood that
the constant zero modes of $\pi$, $\varphi$, $\chi$ and $\xi$ are also
reabsorbed.

v) $d\Phi$ denotes the functional integration over the local degrees
of freedom of the fields. The action $S$ depends on the local degrees
of freedom of the fields, as well as on the moduli $m_i$,
$\nu_j$ and supermoduli $\hat m$ and
$\hat\nu$. The other fields
(Lagrange multipliers, auxiliary fields and some ghost
fields) are those that we have integrated away.

vi) ${\rm e}^{{q_i\over 2}\tilde\pi(z_i)}$ are the $\delta$-type
insertions that simulate the curvature $R$ such that
$\int_{\Sigma_g}R=2(1-g)$. $\tilde\pi$ is the BRST invariant
extension of $\pi$ and the $q_i$ satisfy the condition
\begin{equation}
\sum_i q_i=2(1-g).
\end{equation}
One finds the solution (left moving part)
\begin{equation}
\tilde\pi=\pi+\chi-{2\over \gamma^+}c^z\lambda_- .
\label{tildepi}
\end{equation}

\section{The topological twist on the gauge-fixed theory}
\label{twist1}

In this section, we perform the topological twist on the N=2
gauge-fixed theory. We know that the formal set-up for the topological
twist is entirely encoded into the broker, which correctly changes the
spin, the ghost number and the BRST charge. However, in the gauge-fixed
conformal theory, as we show in a moment, we have more equivalent
possibilities. In particular, we can
adapt the formalism in order to make more evident
contact with the well-known
procedure \cite{eguchiyang} in conformal field theory
that consists of redefining the energy momentum tensor by adding to it
the derivative of the $U(1)$ current \cite{topLG}.
In particular, we conveniently
separate the operation of changing the spin from the rest of the
twist procedure (change of ghost number and BRST charge), the rest still being
performed by the broker.

In order to produce a twisted energy-momentum tensor equal to
$T_{zz}+{1\over 2}\partial_{z} J_z$ we can make
a redefinition of the ghost $c$ of the form
\begin{equation}
c^\prime=c-\partial_{z} c^z.
\label{redef1}
\end{equation}
Such a replacement, which changes the spin of the fields,
is to be viewed as a
redefinition of the $U(1)^\prime$ ghost $C^\prime$ rather
than the $U(1)$ ghost $C$, since the new spin is defined
(see section \ref{twist}) by adding
the $U(1)^\prime$ charge
(not the $U(1)$ charge) to the old spin.
$c^\prime$ has a nonvanishing operator product expansion with
$b_{zz}$ so that it is also necessary to redefine $b_{zz}$, namely
\begin{equation}
b_{zz}^\prime=b_{zz}-\partial_{z} b_{z},
\label{redef2}
\end{equation}
in order to preserve
the operator product expansions.

The spin changes can be read from table \ref{topotable} and justify
the following change in notation
\begin{equation}
\begin{array}{llll}
\eta_z^+\rightarrow\eta_z,\quad &\quad\lambda_+\rightarrow \lambda,
\quad &\quad \beta_{+z}\rightarrow \beta_z,\quad &\quad
\gamma^+\rightarrow \gamma,\\
\eta^-_{z}\rightarrow\eta,\quad &\quad\lambda_-\rightarrow \lambda_z,
\quad &\quad \beta_{-z}\rightarrow \beta_{zz},\quad &\quad
\gamma^-\rightarrow \gamma^z.
\end{array}
\end{equation}
Similarly, the supercurrents become
\begin{equation}
G_{+z}\rightarrow G_z,\quad\quad G_{-z}\rightarrow G_{zz}.
\end{equation}

Redefinitions (\ref{redef1}) and (\ref{redef2}) produce a new
BRST current $J_z^{\prime\, BRST}$ (equal to the old one apart from a
total derivative term) given by
\begin{equation}
J^{\prime\, BRST}_z=-c^z{\cal T}_{zz}^\prime
+{1\over 2}c^\prime {\cal J}_{z}
+{1\over 2}\gamma{\cal G}_{z}-{1\over 2}\gamma^z{\cal G}_{zz},
\end{equation}
where ${\cal T}_{zz}^\prime={\cal T}_{zz}+{1\over 2}\partial_z {\cal J}_z$.
As anticipated,
$J^{\prime\, BRST}_z$ generates
a new energy-momentum tensor (obtained by acting
with the new BRST charge on $b_{zz}^\prime$) equal to
\begin{eqnarray}
T_{zz}^\prime&=&T_{zz}+{1\over 2}\partial_{z} J_z
=-\partial_{z} \pi\partial_{z}\varphi+{1\over 2}\partial_{z}^2\pi+
\partial_{z}\chi\partial_{z}\xi+{1\over 2}\partial_{z}^2 \chi-\lambda_z
\partial_{z}\eta
-\partial_{z}\lambda \eta_z
\nonumber\\&&+2b^\prime_{zz}\partial_{z} c^z+\partial_{z} b_{zz}^\prime c^z
+2\beta_{zz}\partial_{z}\gamma^z+\partial_{z}\beta_{zz}\gamma^z
+\beta_{z}\partial_{z}\gamma+b_{z}\partial_{z} c^\prime.
\end{eqnarray}
{}From this expression, it is immediate to check the new spin assignments.
It is interesting to note that the total derivative term in
the $U(1)$ current $J_z^{gh}$ (\ref{ghrepr})
combines with redefinitions (\ref{redef1})
and (\ref{redef2}) to give the correct energy-momentum tensor
for the ghosts $T_{zz}^{\prime\, gh}$.

The other ingredient of the topological twist is the topological shift
\cite{ansfre}
\begin{equation}
\gamma\rightarrow \gamma+\alpha.
\label{redef3}
\end{equation}
Since the spin has been already changed by
(\ref{redef1}), (\ref{redef3}) does not change the spin
a second time. Indeed, the new spin of
$\gamma$ is zero and so that of $\alpha$.
Moreover, after twist
$\gamma$ possesses a zero mode (the constant).
In this case, $\alpha$
represents a shift of the zero mode of $\gamma$.

(\ref{redef1}), (\ref{redef2}) and (\ref{redef3}) permit to
move directly from the N=2 amplitudes to the amplitudes of the topologically
twisted theory\footnotemark
\footnotetext{However, one has to be careful about the zero modes and
the global degrees of freedom. Later on we shall come back to this point.}.
This shows that the amplitudes of the topological theory are a
subset of the amplitudes of the N=2 supersymmetric theory.
Nevertheless, the twisting procedure cannot be
described simply as a change of variables in the functional integral,
as (\ref{redef1}), (\ref{redef2}) and (\ref{redef3})
should seem to suggest,
since the physical content of the theory could not be changed
in this way.
This is made apparent by the fact that before twist $\gamma$
($\gamma^+$) has spin $-1/2$ and possesses no zero mode, so that
the choice $\alpha=$const is only consistent after redefining the spin
with (\ref{redef1})
and (\ref{redef2}), while in section \ref{twist} it was only
a formal position.

For convenience, as far as the gradings are concerned (ghost number
and fermion number), we use the same conventions as before twist
and leave the broker explicit. Thus $\alpha$ carries odd fermion number
and odd ghost number.

The topological shift (\ref{redef3})
produces a total BRST current equal to
\begin{equation}
J_z^{BRST\, tot}=J_z^{\prime BRST}+{1\over 2}\alpha {G}_{z},
\end{equation}
(again, a total derivative term has been omitted).
If we denote, as usual,
${\cal Q}_{BRST}=\oint J_z^{BRST\, tot}$, ${\cal Q}_v
=\oint J_z^{\prime\, BRST}dz$ and ${\cal Q}_s=\oint G_{z}dz$,
we see that the BRST charge is precisely shifted by the supersymmetry
charge ${\cal Q}_s$,
as explained in section \ref{twist}.

Let us now discuss some properties of the twisted theory.
It is convenient to write down the
${\cal Q}_s$ transformations of the fields.
We denote it by $\delta_s$.
\begin{equation}
\matrix{
\delta_s(\xi-\varphi)=2\eta,&
\delta_s\eta=0,&
\delta_s\lambda_z=\partial_{z}(\pi+\chi), &
\delta_s(\pi+\chi)=0,\cr
\delta_s(\pi-\chi)=2\lambda,&
\delta_s\lambda=0,&
\delta_s\eta_z=\partial_{z} (\xi+\varphi),&
\delta_s(\xi+\varphi)=0,\cr
\delta_s b_{zz}^\prime=0,&
\delta_s \beta_{zz}=-b_{zz}^\prime,&
\delta_s c^z=\gamma^z,&
\delta_s\gamma^z=0,\cr
\delta_s b_{z}=\beta_{z},&
\delta_s\beta_{z}=0,&
\delta_s c^\prime=0,&
\delta_s \gamma=c^\prime.}
\label{ation}
\end{equation}
These transformations are the analogue, in the gauge-fixed case, of
the $\delta_T$ transformations (\ref{deltaT}) and (\ref{deltaT2}).
As explained in the previous section, there are two
$b$-$c$-$\beta$-$\gamma$ systems, rearranged in the
last two lines of (\ref{ation}). In particular, the last line
represents the sector of ${\cal B}_{gauge-fixing}$ that
is reminiscent of the constraint on the moduli space.
The last but one line represents the usual $b$-$c$-$\beta$-$\gamma$
ghost for ghost system
of topological gravity \cite{verlindesquare}.
It is evident
that the roles of $b$ and $\beta$ and the roles of $c$
and $\gamma$ are inverted in the two cases\footnotemark
\footnotetext{In ref.\ \cite{distler},
it is claimed that the topological twist of N=2 supergravity leads
to the Verlinde and Verlinde model. To obtain this, a certain reduction
mechanism in
the ghost sector is advocated,
corresponding to setting $\gamma=0$ and $c^\prime=0$.
The neglected sector is precisely the sector
that is responsible for the constraint on the moduli
space, i.e.\ the last line of (\ref{ation}), which makes the difference
between our model and the model of \cite{verlindesquare}.}.

The theory is topological, since the energy-momentum tensor $T_{zz}^\prime$
is a physically trivial left moving operator.
Indeed, recalling that $G_{zz}=-2\{{\cal Q}_v,
\beta_{zz}\}$, we have
\begin{equation}
\alpha  T_{zz}^\prime=\{{\cal Q}, G_{zz}\}, \quad \quad \{{\cal Q}_v,
G_{zz}\}=0.
\end{equation}

In ref.\ \cite{eguchietal}, it is shown that a ``homotopy'' operator
$U$ can be defined in the Verlinde and Verlinde model of topological
gravity \cite{verlindesquare}, such that $U{\cal Q}_{BRST}U^{-1}={\cal Q}_s$.
This shows that ${\cal Q}_{BRST}$ and ${\cal Q}_s$ have the same spectra
and provides a ``matter'' representation of the gravitational
observables when the theory is coupled to a Landau-Ginzburg model.
We now find the operator $U$ in our case and study some of its
properties. This is not only useful for the future program of coupling
matter to constrained topological gravity, but also permits
to define a third nilpotent operator, called $S$,  that only acts
on the ghost sector and further puts into evidence, in some sense,
the presence of the constraint on moduli space

To begin with, it is straightforward to prove that
\begin{equation}
J_z^{\prime\, BRST}={1\over 2}\delta_s (\gamma{\cal J}_{z}-c^z{\cal G}_{zz}).
\label{eqo1}
\end{equation}
Moreover, one also has
\begin{equation}
{1\over 2}(\gamma{\cal J}_{z}-c^z{\cal G}_{zz})=\delta_v
(\gamma b_{z}-c^z\beta_{zz})+{1\over 4}\delta_s (c^z\gamma\beta_{zz}
+b_{z}\gamma\gamma),
\label{eqo2}
\end{equation}
where $\delta_v$ denotes the action of ${\cal Q}_v$.
Now, defining
\begin{equation}
\Theta={1\over 2}\oint(\gamma{\cal J}_{z}-c^z{\cal G}_{zz})-
{1\over 4}\delta_s \oint (c^z\gamma\beta_{zz}
+b_{z}\gamma\gamma),
\end{equation}
one can write
\begin{equation}
\matrix{
{\cal Q}_v=\{{\cal Q}_s,\Theta\},\quad &\quad
\{{\cal Q}_v,\Theta\}=0.}
\end{equation}
Thus, the desired ``homotopy'' operator is
\begin{equation}
U={\rm exp}\,\left(-2\alpha^{-1}\Theta\right)
\end{equation}
and we have
\begin{equation}
U{\cal Q}U^{-1}={1\over 2}\alpha{\cal Q}_s.
\end{equation}
In this way, it is possible to turn to the ``matter picture'', which
is also simpler from the computational point of view.
The key point, in presence of matter coupling, is that
the condition of equivariant cohomology is correspondingly
changed \cite{eguchietal}.

Let us now introduce the operator $S$.
We have, from (\ref{eqo1}) and (\ref{eqo2}),
\begin{equation}
J_z^{\prime\, BRST}=\delta_v\delta_s(\gamma b_{z}-c^z\beta_{zz}).
\label{8.16}
\end{equation}
Due to $\delta_v J_z=0$,
we can also write
\begin{equation}
[{\cal Q}_{gh}^\prime,{\cal Q}_v]={\cal Q}_v,
\end{equation}
where
\begin{equation}
{\cal Q}_{gh}^\prime={\cal Q}_{gh}+\oint J_z=
\oint b_{zz}^\prime c^z+2\beta_{zz}\gamma^z+b_zc^\prime-\lambda_z
\eta
-\lambda\eta_z
\end{equation}
is the ghost number charge of the twisted theory, equal to the sum
of the ghost number charge of the initial N=2 theory plus the $U(1)$
charge. This corresponds to eq.\ (\ref{spina}).

Define
\begin{equation}
S=\oint(\gamma
b_{z}-c^z\beta_{zz}).
\end{equation}
Then, eq.\ (\ref{8.16}) implies
\begin{equation}
{\cal Q}_v=[{\cal Q}_v,\{{\cal Q}_s,S\}],\quad\quad
\Theta=[{\cal Q}_v, S],
\end{equation}
while the ghost charge can be expressed as
\begin{equation}
{\cal Q}^\prime_{gh}=\oint J_z-\{{\cal Q}_s,S\}.
\end{equation}
$S$ is a nilpotent operator, that acts trivially on the Liouville
sector, while in the ghost sector it gives
\begin{equation}
\matrix{Sb_{zz}^\prime=-\beta_{zz},&
S\beta_{zz}=0,&
S c^z=0,&
S\gamma^z=-c^z,\cr
S b_{z}=0,&
S \beta_{z}=b_z,&
S c^\prime=-\gamma,&
S \gamma=0.}
\end{equation}
These rules should be compared with the last two lines of (\ref{ation}).
In some sense,
the action of $S$ is dual to the action of $\delta_s$ in the ghost sector.
We have already discussed the last two lines of (\ref{ation})
and the difference between the roles of $b$ and $\beta$,
$c$ and $\gamma$, in the two cases. The action of the operator $S$,
compared to the one of $\delta_s$, inverts the roles of the two
$b$-$c$-$\beta$-$\gamma$ systems.
Clearly, the existence of the operator $S$
is strictly related to the presence of the
graviphoton and thus to the constraint on moduli space.

Let us give some of the transformations corresponding to the change
of basis due to $U$
\begin{equation}
\begin{array}{cc}
U\pi U^{-1}=\pi+{1\over 1-\gamma/\alpha}{c^z\lambda_z\over \alpha},&\quad
U\varphi U^{-1}=\varphi+{1\over 1-\gamma/\alpha}{c^z\eta_z\over \alpha},\\
U\chi U^{-1}=\chi+{1\over 1-\gamma/\alpha}{c^z\lambda_z\over \alpha},&\quad
U\xi U^{-1}=\xi+{1\over 1-\gamma/\alpha}{c^z\eta_z\over \alpha}+
\ln(1-\gamma/\alpha),\\
U\lambda_z U^{-1}={1\over 1-\gamma/\alpha}\lambda_z,&\quad
U\eta U^{-1}=(1-\gamma/\alpha)\eta+f(\varphi,\xi,\gamma,\eta_z,c^z),\\
U\eta_z U^{-1}={1\over 1-\gamma/\alpha}\eta_z,&\quad
U\lambda U^{-1}=(1-\gamma/\alpha)\lambda+f(\pi,\chi,\gamma,
\lambda_z,c^z),\\
Uc^zU^{-1}={1\over 1-\gamma/\alpha}c^z,&\quad
Ub_{zz}^\prime U^{-1}=b_{zz}^\prime-{1\over \alpha} G_{zz},\\
Uc^\prime U^{-1}={1\over 1-\gamma/\alpha}c^\prime+f(\gamma,c^z),&\quad
Ub_zU^{-1}=b_{z}+{1\over \alpha} (c^z\beta_{zz}-\gamma b_z),\\
U\gamma^z U^{-1}=\gamma^z+f(\gamma,c^z,c^\prime),&\quad
U\beta_{zz}U^{-1}=\beta_{zz},\\
U\gamma U^{-1}={1\over 1-\gamma/\alpha}\gamma,&\quad
U\beta_{z}U^{-1}=(1-\gamma/\alpha)^2
\beta_{z}+f({\rm all\, but}\,\beta_z),\\
UT_{zz}^\prime U^{-1}=T_{zz}^\prime,&\quad
UJ_zU^{-1}=J_z-{2\over \alpha}\theta_z,\\
UG_{zz}U^{-1}=G_{zz},&\quad
UG_{z}U^{-1}=G_{z}-{2\over \alpha}\delta_s\theta_{z},
\end{array}
\label{arra}
\end{equation}
where $\theta_z$ is such that
$\Theta=\oint \theta_zdz$.
$T_{zz}^\prime$, $G_{z}-{2\over \alpha}\delta_s\theta_z$,
$G_{zz}$ and $J_z-{1\over 2\alpha}\theta_z$ is another
representation of the same topological algebra.
The functions $f$ appearing in (\ref{arra}) are complicated
expressions of their arguments, which we do not report here.
The above information is sufficient to prove
that the operator $U$ defines a changes of variables in the functional
integral with unit Jacobian determinant.

Notice that $\pi+\chi$ is the field that permits the insertion
of curvature delta-type singularities. It is ${\cal Q}_s$-closed
and $\tilde\pi=U(\pi+\chi)U^{-1}$ is its ${\cal Q}$-closed generalization.
Moreover, since $UT^\prime_{zz}U^{-1}=T^\prime_{zz}$, it is immediate
to prove that $\tilde\pi$ has the same operator product expansion
with $T^\prime_{zz}$ as
$\pi$. In particular, ${\rm e}^{\tilde\pi}$ is a primary field. The limit
$\alpha\rightarrow 0$ of $\tilde\pi$ is the field (\ref{tildepi})
that allowed the curvature insertions in the N=2 theory (also
called $\tilde\pi$).

\section{Geometrical Interpretation}
\label{geometry}

We now discuss the moduli space of the twisted theory
and the gauge-fixing sector that implements
the constraint defining the submanifold ${\cal V}_g\subset {\cal M}_g$.

The number of moduli of the twisted theory is $4g-3$,
the same as that of
the N=2 theory, $3(g-1)$ moduli $ m_i $ corresponding to the metric
and $g$ moduli $\nu_j$
corresponding to the $U(1)$ connection $A$.
The number of supermoduli,
on the other hand, changes by one: it was $4(g-1)$ for the N=2
theory, it is $4g-3$ for the topological theory, $3(g-1)$ supermoduli
$\hat  m_i $
corresponding
to the zero modes of the spin 2 antighost $\beta_{zz}$ and $g$
supermoduli $\hat \nu_j$
corresponding to the zero modes of $\beta_{z}$.
The mismatch of one supermodulus is filled by the presence of
one super Killing vector field, corresponding to the (constant)
zero mode of $\gamma$.

Thus, comparing the N=2 theory with the twisted one, we can say that
the $2(g-1)+2(g-1)$ zero modes of the $\beta_{\pm z}$ fields rearrange into
the $3(g-1)$  zero modes of $\beta_{zz}$ plus the $g$ zero modes of $\beta_z$
plus one zero mode of $\gamma$.
Similarly, the $2(g-1)+2(g-1)$ supermoduli rearrange into
$3(g-1)$ moduli $\hat m$ of the topological ghosts plus
$g$ moduli $\hat \nu$
of the topological antighosts plus one super Killing vector field.
The zero modes of the $\lambda$ and $\eta$ fields rearrange
among themselves.

In particular, after the twist,
the number of bosonic moduli equals the number of fermionic moduli,
as expected for a topological theory.
However, the two kinds of supermoduli $\hat  m $ and $\hat \nu$
do not carry the same ghost number
after the twist. Indeed, $\hat  m_i $ carry ghost number $1$, while
$\hat \nu_j$ carry ghost number $-1$.
Thus, we can interpret $\hat  m_i $ as the topological variation
of $ m_i $, but we cannot interpret $\hat \nu_j$ as the topological
variation of $\nu_j$, rather $\nu_j$ is the topological
variation of $\hat \nu_j$:
\begin{equation}
\matrix{
s m_i =\hat  m_i ,&\quad s\hat  m_i =0,&
\quad s\hat \nu_j=\nu_j,&\quad s \nu_j=0.}
\label{smu}
\end{equation}
This is in agreement with the interpretation of $A$ as a Lagrange
multiplier, so that it is only introduced {\sl via} the gauge-fixing algebra:
$ m $ and $\hat  m $ belong to ${\cal B}_{gauge-free}$, while
$\nu$ and $\hat \nu$ belong to ${\cal B}_{gauge-fixing}$.

The amplitudes can be written as
\begin{equation}
<\prod_k\sigma_{n_k}>=\int d\Phi
\int_{{\cal M}_g}\prod_{i=1}^{3g-3}d m_i
\int_{{{\bf C}^g / \Lambda}}\prod_{j=1}^gd\nu_j
\int d\hat m  d\hat\nu
\,\prod_i{\rm e}^{q_i\tilde\pi(z_i)}\,
{\rm e}^{-S( m ,\hat m ,\nu,\hat\nu)}\prod_k\sigma_{n_k},
\label{ampltw}
\end{equation}
where $\sigma_{n_k}$ are the observables.
In this expression, the insertions that remove the zero modes of
$b_{zz}$, $\beta_{zz}$, $\beta_{z}$, $b_{z}$, $\eta$, $\lambda$, $\lambda_z$
and $\eta_z$
are understood, but attention has to be paid to the fact that a
super Killing vector field, corresponding to the zero mode of $\gamma$,
forbids one fermionic integration.
The ghost number of the supermoduli measure
adds up to $-2g+3$. Nevertheless, due to the presence
of one super Killing vector field,
the selection rule is that the total ghost number
of $\prod_k\sigma_{n_k}$ must be equal to $2(g-1)$ and not
to $2g-3$. This is the mismatch between true dimension and formal
dimension addressed in the introduction.

To explain why the graviphoton is responsible for the
constraint, let us rewrite the action making the dependence
on the $U(1)$-moduli $\nu_j$ and the corresponding
supermoduli $\hat \nu_j$ explicit.
\begin{eqnarray}
S( m ,\hat m ,\nu,\hat\nu)&=&S( m ,\hat m ,0,0)
+\nu_j\int_{\Sigma_g}\omega^{j}_{\bar z}J_{z}d^2z+
\hat\nu_j\int_{\Sigma_g}\omega^{j}_{\bar z}G_{z}d^2z\nonumber\\&&+
\bar\nu_j\int_{\Sigma_g}\omega^{j}_{z}J_{\bar z}d^2z+
\hat{\bar\nu}_j\int_{\Sigma_g}\omega^{j}_{z}G_{\bar z}d^2z
+\nu\-\hat\nu{\rm -terms}.
\label{actio}
\end{eqnarray}
The terms that are quadratic in $\nu\-\hat\nu$ are due
to the fact that the gravitini are initially $U(1)$-charged. They have
not been reported explicitly, since they can be neglected, as we show
in a moment.
The coefficient of $\bar\nu_j$ is the $U(1)$ current $J_z$
folded with the $j$-th (anti)holomorphic differential $\omega^j_{\bar z}$.
Similarly, the coefficient of ${\hat {\bar\nu}}_j$ is the supercurrent
$G_z$ folded with the same differential.

We want to perform the $\nu$-$\hat\nu$ integrals explicitly.
This is allowed, since the observables should not depend
on  $\nu$ and $\hat\nu$. Indeed,  $\nu$ and $\hat\nu$
belong to ${\cal B}_{gauge-fixing}$ and not to ${\cal B}_{gauge-free}$,
while the observables are constructed entirely
from ${\cal B}_{gauge-free}$. Anyway, since $\nu$ and $\hat\nu$
form a closed BRST subsystem, we can consistently
project down to the subset
$\nu=\hat\nu=0$, while retaining the BRST nilpotence.
The $U(1)$ moduli $\nu$ are not integrated all over ${\bf C}^g$,
which would be nice since the integration would be very easy, rather
on the unit cell $L={\bf C}^g/({\bf Z}^g+\Omega {\bf Z}^g)$
defined by the Jacobian lattice. To overcome this problem,
we take the semiclassical limit, which is exact in a topological
field theory. We multiply the action $S$ by a constant $\kappa$
that has to be stretched to infinity. $\kappa$ can
be viewed as a gauge-fixing
parameter, rescaling the gauge-fermion: no physical amplitude depends
on it.
Let us define
\begin{equation}
\nu^\prime_j=\kappa\nu_j\quad\quad \hat \nu^\prime_j=\kappa\hat\nu_j.
\end{equation}
We have
\begin{equation}
\int_L\prod_{j=1}^gd\nu_jd\hat\nu_j=
\int_{\kappa L}\prod_{j=1}^gd\nu^\prime_jd\hat\nu_j,
\label{p1}
\end{equation}
where  and $\kappa L$ is unit cell rescaled.
We see that the $\nu$-$\hat\nu$-terms of (\ref{actio}) are
suppressed in the $\kappa\rightarrow\infty$ limit, as claimed.
We can replace $\kappa L$ with ${\bf C}^g$ in this limit.
Finally, the integration over the $U(1)$ moduli and supermoduli
produces the insertions
\begin{equation}
\prod_{j=1}^g\int_{\Sigma_g} \omega^j_{\bar z}G_zd^2z\cdot\delta\left(
\int_{\Sigma_g}\omega^j_{\bar z}J_zd^2z\right).
\label{2.47}
\end{equation}

The delta-function is the origin of the desired constraint on moduli space.
Indeed, the current $J_z$ can be thought as a (field dependent)
section of ${\cal E}_{hol}$. The requirement of its vanishing
is equivalent to projecting onto the Poincar\'e dual of the top Chern
class $c_g({\cal E}_{hol})$
of ${\cal E}_{hol}$, due to a theorem that one can find for example in
\cite{griffithsharris}. Changing section only changes the representative
in the cohomology class of $c_g({\cal E}_{hol})$.
Indeed, the Poincar\`e dual of the top Chern class of a holomorphic
vector bundle $E\rightarrow M$ is shown to be the submanifold of the base
manifold $M$ where one holomorphic section $a\in \Gamma(E,M)$
vanishes identically. In other words, the dual of $c_{g}({\cal
E}_{hol})$ is the divisor of some section. For a line bundle
$L\rightarrow M$, this is easily seen. Let $h$ be a fiber metric so
that
$||a||^2=a(z)\bar a(\bar z)h(z,\bar z)$ is the norm of the
section $a$. The top Chern class $c_1(L)$ can be written as the
cohomology class of the curvature $R=\bar \partial\Gamma$
of the canonical holomorphic connection $\Gamma=h^{-1}\partial h$,
so that
\begin{equation}
c_1(L)=\bar \partial\partial\ln ||a(z)||^2.
\end{equation}
Patchwise, the metric $h$ can be reduced to the identity, but then
$c_1(L)$ becomes a de Rahm current, namely a singular $(1,1)$ form
with delta-function support on the divisor Div[$a$], i.e.\ the
locus of zeroes and poles of $a(z)$. The divisor Div[$a$]
is the Poincar\`e dual of $c_1(L)$.
For a holomorphic vector bundle $E\rightarrow M$ of rank $n$, the
same theorem can be understood using the so-called splitting principle.
For the purpose of calculating Chern classes, $E$ can always be
regarded
as the Whitney sum of $n$ line-bundles $L_i$ corresponding,
naively, to the eigendirections of the curvature matrix two form
${\cal R}^{jk}$,
\begin{equation}
E=L_1\oplus\cdots\oplus L_n.
\end{equation}
Then we have
\begin{equation}c_n(E)=\prod_{i=1}^nc_1(L_i)=
\bar\partial\partial\ln ||a_1||^2\wedge\cdots\wedge
\bar \partial\partial\ln ||a_n||^2,
\end{equation}
where $a_i$ are the components of a section $a$ in a suitable basis.
{}From this formula, we see that $c_n(E)$ has delta-function support
on the divisor of $a$.
That is why in our derivation of the topological correlators from the
functional integral, we do not pay particular
attention to the explicit form of $J_z$ and to its dependence on the
other fields. What matters is that it is a conserved holomorphic one
form, namely a section of ${\cal E}_{hol}$. The functional integral
imposes its vanishing, so that the Riemann surfaces
that effectively contribute lie in the homology class of the
Poincar\`e dual of $c_g({\cal E}_{hol})$.

Summarizing, we argue that the topological observables $\sigma_{n_k}$
correspond to the Mumford-Morita classes, as in the
case of topological gravity \cite{verlindesquare},
but that in constrained topological gravity the
correlation functions are intersection forms
on the Poincar\'e dual ${\cal V}_g$ of $c_g({\cal E}_{hol})$ and not on the
whole moduli space ${\cal M}_g$.

It can be convenient to
represent
$c_g({\cal E}_{hol})$ by introducing the natural
fiber metric $h_{jk}={\rm Im}\, \Omega^{jk}=\int_{\Sigma_g}
\omega^j_z\omega^k_{\bar z}d^2z$ on ${\cal E}_{hol}$.
The canonical connection associated
with this metric is then
\begin{equation}
\Gamma=h^{-1}\partial h={1\over \Omega-\bar \Omega}\partial \Omega,
\end{equation}
which leads to a curvature
\begin{equation}
{\cal R}=\bar\partial\Gamma=
{1\over \Omega-\bar \Omega}\bar\partial\bar\Omega
{1\over \Omega-\bar \Omega}\partial\Omega.
\end{equation}

Let $\{\omega^1,\ldots \omega^g\}$
denote a basis of holomorphic differentials.
Locally, we can expand $J_z$ in this basis
\begin{equation}
J_z=a_j\omega^j_z.
\end{equation}
The field dependent
coefficients $a_j$ are the components of the section $J_z\in\Gamma(
{\cal E}_{hol},{\cal M}_g)$.
The constraint then reads
\begin{equation}
{\rm Im}\,\Omega^{jk}a_k=0,\quad\quad\forall j,
\end{equation}
which, due to the positive definiteness of ${\rm Im}\,\Omega$, is equivalent
to
\begin{equation}
a_j=0,\quad\quad\forall j.
\end{equation}
These are the equations that (locally) identify the
submanifold ${\cal V}_g\subset {\cal M}_g$.
It is also useful to introduce
the vectors $v_j={\partial\over \partial a_j}$ that
provide a local basis for the normal bundle ${\cal N}({\cal V}_g)$ to
${\cal V}_g$. Of course, the vectors $v_j$ commute among themselves:
\begin{equation}
[v_j,v_k]=0.
\end{equation}
In these explicit local coordinates, the top Chern class
$c_g({\cal E}_{hol})$ admits the following representation as a
de Rham current:
\begin{equation}
c_g({\cal E}_{hol})=\delta({\cal V}_g)\tilde\Omega_g,
\label{9.15}
\end{equation}
where
\begin{equation}
\tilde\Omega_g=\prod_{j=1}^g da_j,\quad\quad
\delta({\cal V}_g)=\prod_{j=1}^g\delta (a_j).
\end{equation}
This explicit notation is useful to trace back the correspondence
between the geometrical and field theoretical definition of the
correlators.

To begin with,
a convenient representation of the BRST operator (\ref{smu})
on the space
$\{ m ,\hat  m ,\nu,\hat\nu\}$ is given by
\begin{equation}
{\cal Q}_{global}=\hat  m_i {\partial\over \partial  m_i }+
\nu_j{\partial\over \partial \hat \nu_j}.
\label{operator}
\end{equation}
${\cal Q}_{global}$
is not the total BRST charge, rather it only represents the
BRST charge on the sector of the
global degrees of freedom. The total BRST
charge is the sum of the above operator plus the usual BRST charge
${\cal Q}={\cal Q}_s+{\cal Q}_v$,
that acts only on the local degrees of freedom.
Since the total BRST charge acts trivially inside the physical
correlation functions, we see that the action of ${\cal Q}$
inside correlation
functions is the opposite of the action of ${\cal Q}_{global}$.
This means that
${\cal Q}$ can be identified, apart from an overall immaterial sign,
with the operator (\ref{operator}).
We know that the geometrical meaning of the supermoduli $\hat m_i$
are the differentials $d m_i $ on the moduli space ${\cal M}_g$
and that
the ghost number corresponds to the form degree. In view of this, we
argue that the geometrical meaning of the $U(1)$ supermoduli
$\hat \nu_j$ are contraction operators
$i_{v_j}$ with respect to the associated vectors $v_j$.
Since the $U(1)$ moduli $\nu_j$ are the BRST variations of $\hat\nu_j$
and the BRST operation should be
identified with the exterior derivative,
it is natural to conjecture that $\nu_j$ correspond to
the Lie derivatives along the vectors $v_j$.

The correspondence between field theory and geometry is summarized
in table \ref{table1}.
We now give arguments in support of this interpretation.

For instance, since ${\cal Q}\sim d$,
${\rm Im}\,\Omega^{jk}\, a_k\sim \int\omega_{\bar z}^kJ_zd^2z$ and
$[{\cal Q},J_z]=-G_z$, then
the insertions $\int \omega_{\bar z}^jG_zd^2z$ correspond to
$d({\rm Im}\,\Omega^{jk}a_k)$, so that
\begin{equation}
\prod_{j=1}^g\int_{\Sigma_g} \omega^j_{\bar z}G_zd^2z\cdot\delta\left(
\int_{\Sigma_g}\omega^j_{\bar z}J_zd^2z\right)\sim
\tilde\Omega_g\delta({\cal V}_g)=c_g({\cal E}_{hol}).
\end{equation}

If $\alpha_k$ denote the Mumford-Morita classes corresponding
to the observables ${\cal O}_k$, the amplitudes are
\begin{equation}
<{\cal O}_1\cdots {\cal O}_n>=
\int_{{\cal M}_g}\delta({\cal V}_g)\tilde\Omega_g\wedge \alpha_1
\wedge\cdots\wedge \alpha_n=\int_{{\cal V}_g}\alpha_1
\wedge\cdots\wedge \alpha_n.
\end{equation}

{}From the geometrical point of view, it is immediate to
show that the action of (\ref{operator}) on a correlation
function is precisely the exterior derivative, as already advocated.
Indeed, we can write the $d$-form $\omega_d$ corresponding to a
physical amplitude (not necessarily a top form, if we freeze, for the moment,
the integration over the global degrees of freedom) as
\begin{equation}
\omega_d=i_{v_1}\cdots i_{v_g}\Omega_{d+g}=
\left(\prod_{j=1}^g\hat\nu_j\right)\hat m_{i_1}\cdots\hat m_{i_d}
\Omega_{d+g}^{i_1\cdots i_d},
\end{equation}
where $\Omega_{d+g}$ is a suitable $d+g$-form on ${\cal M}_g$
(equal to $\tilde\Omega_g\wedge\omega_d$).
Now, using the representation (\ref{operator}) of the operator
${\cal Q}$, we find
\begin{eqnarray}
\{{\cal Q},\omega_d\}&=&(-1)^g\left(\prod_{j=1}^g\hat\nu_j\right)
\hat m_i \hat m_{i_1}\cdots\hat m_{i_d}
{\partial\Omega_{d+g}^{i_1\cdots i_d}\over\partial  m_i }\nonumber\\&&+
\sum_{k=1}^g(-1)^{k+1}\nu_k\left(\prod_{j\neq k}\hat\nu_j\right)
\hat m_{i_1}\cdots\hat m_{i_d}
\Omega_{d+g}^{i_1\cdots i_d}.
\end{eqnarray}
Using the correspondence given in table \ref{table1}, we have
\begin{equation}
\{{\cal Q},\omega_d\}=(-1)^gi_{v_1}\cdots i_{v_g}d\Omega_{d+g}
+\sum_{k=1}^g(-1)^{k+1}{\cal L}_{v_k}
i_{v_1}\cdots i_{v_{k-1}}i_{v_{k+1}}\cdots i_{v_g}\Omega_{d+g}=
d\omega_d.
\end{equation}
The second piece of (\ref{operator}) replaces a
contraction with the vector
$v_j$ with the Lie derivative with respect to the same vector.

Finally, we describe a more intuitive description of the
submanifold ${{\cal V}_g}$ \footnotemark\footnotetext{
We would like to thank C.\ Reina for essential private
discussions on this point.}.
${{\cal V}_g}$ is a representative of a homology cycle, so it can be
convenient to make a special choice of this representative,
for example, a ${{\cal V}_g}$ lying on the boundary of the moduli space.
That means that we are considering degenerate Riemann surfaces.
Take $g$ independent and non intersecting homology
cycles on the Riemann surface, the $A$-cycles,
and pinch them. You get $g$ nodes. Then,
separate the two branches of each node: you get a sphere $S_{2g}$
with $2g$ pairwise identified marked points. We conjecture
that ${{\cal V}_g}$ is representable as the space of such spheres.
The dimensions turn out to match: indeed, $2g$ complex
parameters are the positions of the marked points, but, due to
$SL(2,{\bf C})$ invariance on the sphere, three points can be fixed
to $0$, $1$ and $\infty$, as usual. Thus, the dimension of $S_{2g}$
equals $2g-3$, which is the correct result. The $g$ holomorphic
differentials of $\Sigma_g$ become differentials of the third
kind on $S_{2g}$, with opposite residues on the pairwise
identified points.

\section{Outlook and Open Questions}
\label{concl}

In the present paper we have shown that there exists a new class of
topological field theories whose correlation functions can be interpreted
as intersection numbers of cohomology classes in a constrained  moduli space.
The constrained moduli space is, by itself, a homology class of cycles in
an ordinary, unconstrained, moduli space. The specific example that
we have considered is a formulation of 2D topological gravity that is
obtained through the A-twist of N=2 Liouville theory.

Usually, in a topological field theory the space of field
configurations is projected onto a finite dimensional subspace made of
instantons. In two dimensional gravity, however,
the space of configurations is
the moduli space ${\cal M}_g$ of Riemann surfaces of genus $g$ and
ordinary topological gravity deals with the full ${\cal M}_g$.
Constrained topological gravity,
instead, deals with a proper submanifold ${\cal V}_g\subset{\cal M}_g$,
so that also in this case we have a nontrivial concept of instanton
configurations. The ``instantons'' are the solutions to the moduli
space constraint. Our formulation of topological gravity bears the
same relation with 2D gravity as a generic topological field theory
with its non topological version, namely the former is a ``proper''
projection of the latter.

Our result raises several questions that are so far unanswered.

In Witten's topological gravity, where the
correlators are the intersection numbers of Mumford-Morita classes in
the ordinary moduli space ${\cal M}_{g,s}$,
the generating function satisfies
the integrable KdV hierarchy \cite{lez}. This result was shown
by Kontsevich \cite{Kontsevich}
from  algebraic geometry, through a systematic
triangulation of moduli space leading to an integral \'a la matrix-model.
It was also justified, in field theoretical terms,
by the  work of Verlinde and Verlinde \cite{verlindesquare}
and Dijkgraf, Verlinde and Verlinde \cite{deigraf}.
It is clear that the first question
raised by our paper is: {\it which integrable hierarchy is
satisfied by the correlators defined in eq.(\ref{intro_0})?}
The  answer to this question is left open by our work and it is not yet clear
whether it can be more easily obtained from field-theoretical or geometrical
considerations: both ways are equally good, since we have established
a correspondence between the topological definition (\ref{intro_0}) and
the field-theoretical one (\ref{ampltw}).

The next open question concerns matter coupling.
One should investigate the effects of the
moduli space constraint when topological gravity is coupled to topological
matter. This involves the study of the topological twist of matter coupled
N=2 Liouville theory. In the present paper we have extended to curved
superspace the construction of \cite{billofre} where N=2 matter coupled to
gauge theories was analysed. The next step of the program is
to write down the most general N=2, D=2 theory that
contains the graviton multiplet, the gauge multiplets and the chiral and
twisted chiral multiplets \cite{kounnas},
interacting through a  generalized K\"ahler metric,
a superpotential and a dilaton coupling to the two dimensional curvature. Then,
by investigating the A-twist of such a theory, one obtains the matter coupling
of the present {\it constrained topological gravity} to the topological
$\sigma$-model.

If we are interested in coupling the
topological Landau-Ginzburg model, we should instead perform
the B-twist. Indeed, another question
raised by our paper that should be addressed in the next future is
{\it whether the B-twist of the N=2 Liouville theory produces a similar or
different topological gravity}.
In this paper we have shown that the
cohomology of ${\cal Q}_s$ is equivalent to the cohomology
of the full BRST operator ${\cal Q}_{BRST}$, the same way as it happens
for the Verlinde and Verlinde theory. This property can be exploited
fruitfully by coupling the (B-twisted)
topological gravity to Landau-Ginzburg matter.

Finally, other open questions are related to conformal
field theory. The splitting into a minimal plus a maximal model
of the N=2 superconformal theory associated
with the gauge-fixed Liouville model deserves attention. It might be
the way to understand better the relation with the Polyakov formulation
in terms of a level $k$ $SL(2,R)$ Ka\v c-Moody algebra \cite{poly}.

\vspace{24pt}
\begin{center}
{\bf Acknowledgements}
\end{center}

\vspace{12pt}

We are very grateful to our friends B.\ Dubrovin, C.\ Reina, C.\ Becchi,
C.\ Imbimbo, L.\ Bonora, F.\ Gliozzi, G.\ Falqui, M.\ Martellini
and A.\ Zaffaroni
for enlightening and essential discussions.

\begin{table}
\begin{center}
\caption{\sl Twist A}
\begin{tabular}{|c||c|c||c|c||c|c||c|}
\hline
Field   & spin  & ghost  & spin$^\prime$ & ghost$^\prime$ & $U(1)$ &
$U(1)^\prime $ & New Field
\\
\hline\hline
$e^+$ & $-1$ & $0$ & $-1$ & $0$ & $0$ & $0$ & $~~$ \\
\hline
$e^-$ &$1$ & $0$ & $1$ & $0$ & $0$ & $0$ & $~~$\\
\hline
\hline
$C^+$ & $-1$ & $1$ & $-1$ & $1$ &$0$ &$0$ & $~~$\\
\hline
$C^-$ & $1$ &$1$ &$1$ &$1$ &$0$ & $0$ & $~~$\\
\hline
\hline
$\omega$ & $0$ & $0$ &$0$ &$0$ &$0$ &$0$ & $~~$\\
\hline
$C^0$ & $0$ & $1$ &$0$ &$1$ &$0$ &$0$ & $~~$\\
\hline\hline
$A$ & $0$ &$0$ &$0$ &$0$ &$0$ &$0$ & $~~$\\
\hline
$C$ & $0$ &$1$ &$0$ &$1$ &$0$ &$0$ & $~~$\\
\hline\hline
$\zeta^+$ & $- 1/2 $ & $0$ & $0$ &$-1$ &$- 1/2 $ &$ 1/2 $ &
$\zeta^+\alpha^{-1}\equiv \bar\xi$\\
\hline
$\zeta^-$ & $- 1/2 $ & $0$ & $-1$ &$1$ &$ 1/2 $ &$- 1/2 $ &
$\zeta^- \alpha$\\
\hline
$\tilde\zeta_+$ & $ 1/2 $ & $0$ & $1$ &$1$ &$ 1/2 $ &$ 1/2 $
& $\tilde\zeta_+ \beta$\\
\hline
$\tilde \zeta_-$ & $ 1/2 $ & $0$ & $0$ &$-1$ &$- 1/2 $ &$- 1/2 $
& $\tilde\zeta_- \beta^{-1}\equiv \bar {\tilde \xi}$\\
\hline
\hline
$\Gamma^+$ &  $- 1/2 $ & $1$ & $0$ &$0$ &$- 1/2 $ &$ 1/2 $
& $\Gamma^+\alpha^{-1}$\\
\hline
$\Gamma^-$ &  $- 1/2 $ & $1$ & $-1$ &$2$ &$ 1/2 $ &$- 1/2 $
& $\Gamma^-\alpha$\\
\hline
$\tilde\Gamma_+$ &  $ 1/2 $ & $1$ & $1$ &$2$ &$ 1/2 $ &$ 1/2 $
& $\tilde\Gamma_+\beta$\\
\hline
$\tilde\Gamma_-$ &  $ 1/2 $ & $1$ & $0$ &$0$ &$- 1/2 $ &$- 1/2 $
& $\tilde\Gamma_-\beta^{-1}$\\
\hline
\hline
$M$ &  $0$ & $0$ & $1$ &$0$ &$0$ &$1$ & $M\alpha^{-1}\beta=M_+$\\
\hline
$\bar M$ &  $0$ & $0$ & $-1$ &$0$ &$0$ &$-1$ & $
\bar M \alpha\beta^{-1}=M_-$\\
\hline
\hline
$X^I$ & $0$ & $0$ & $0$ & $0$ & $0$ & $0$ & $~~$\\
\hline
$ X^{I*}$ &$0$ & $0$ & $0$ & $0$ & $0$ & $0$ & $~~$\\
\hline
\hline
$\lambda_-, \psi^i$ & $ 1/2 $ & $0$ & $1$ &$-1$ &$- 1/2 $ &$ 1/2 $
& $\lambda_- \alpha^{-1}\equiv\chi_+$\\
\hline
$\lambda_+,\psi^{i*}$ & $ 1/2 $ & $0$ & $0$ &$1$ &$ 1/2 $ &$- 1/2 $
& $\lambda_+ \alpha$\\
\hline
$\tilde\lambda^-,\tilde \psi^i$ & $- 1/2 $ & $0$ & $0$ &$1$ &$ 1/2 $ &$ 1/2 $
& $\tilde\lambda^- \beta$\\
\hline
$\tilde\lambda^+,\tilde\psi^{i*}$ & $- 1/2 $ & $0$ & $-1$ &$-1$ &$- 1/2 $
&$- 1/2 $ &$\tilde\lambda^+ \beta^{-1}\equiv \chi_-$\\
\hline\hline
$H^I$ &  $0$ & $0$ & $1$ &$0$ &$0$ &$1$ &$H\alpha^{-1}\beta$\\
\hline
$H^{I*}$ &  $0$ & $0$ & $-1$ &$0$ &$0$ &$-1$ &$\bar H \alpha
\beta^{-1}$\\
\hline\hline
$\alpha$ &  $- 1/2 $ & $1$ & $0$ &$0$ &$- 1/2 $ &$ 1/2 $
& $~~$\\
\hline
$\beta$ &  $ 1/2 $ & $1$ & $0$ &$0$ &$- 1/2 $ &$- 1/2 $ & $~~$\\
\hline
\end{tabular}
\end{center}
\label{topotable}
\end{table}

\begin{table}
\begin{center}
\caption{Field Theory versus Geometry}
\begin{tabular}{|c|c|}
\hline
Field Theory & Geometry\quad \\
\hline
\quad $\hat  m_i $     & $d m_i $\quad \\
\quad $\hat \nu_j$     & $i_{v_j}$\quad \\
\quad $\nu_j$          & ${\cal L}_{v_j}$\quad \\
\quad ${\cal Q}$       & $d$\quad \\
\quad $[{\cal Q}, m_i ]=\hat m_i$    & $[d, m_i ]=d m_i $\quad \\
\quad $\{{\cal Q},\hat m_i \}=0$      & $\{d,d m_i \}=0$\quad \\
\quad $\{{\cal Q},\hat \nu_j\}=\nu_j$ & $\{d,i_{v_j}\}={\cal L}_{v_j}$\quad \\
\quad $[{\cal Q},\nu_j]=0$            & $[d,{\cal L}_{v_j}]=0$\quad \\
\quad $\{\hat\nu_j,\hat\nu_k\}=0$     & $\{i_{v_j},i_{v_k}\}=0$\quad \\
\quad $[\hat\nu_j,\nu_k]=0$           & $[i_{v_j},{\cal L}_{v_k}]=0$\quad \\
\quad $[\nu_j,\nu_k]=0$               & $[{\cal L}_{v_j},{\cal L}_{v_k}]=0$
\quad \\
\quad $\prod_{j=1}^g \delta\left(\int \omega^j_{\bar z}
J_zd^2z\right)$&
$\delta({{\cal V}_g})$\quad\\
\quad $\prod_{j=1}^g \int\omega^j_{\bar z}
G_zd^2z$ &$\tilde\Omega_g$\quad\\
\quad  $\prod_{j=1}^g \int\omega^j_{\bar z}
G_zd^2z\cdot\delta\left(\int \omega^j_{\bar z}J_zd^2z\right)$
& $c_g({\cal E}_{hol})=\delta({{\cal V}_g})\tilde\Omega_g$\quad\\
\quad $\sigma_{n_j}$   & $[c_1({\cal L}_j)]^{n_j}$\quad\\
\quad $<\sigma_{n_1}\cdots \sigma_{n_k}>$ & $\int_{{\cal V}_g}
[c_1({\cal L}_1)]^{n_1}\wedge\cdots \wedge
[c_1({\cal L}_k)]^{n_k}$\quad \\
\hline
\end{tabular}
\end{center}
\label{table1}
\end{table}

\end{document}